\documentclass[openacc]{rsproca_new_hacked_arxiv}

\usepackage{color}
\usepackage{bm}
\usepackage[figuresright]{rotating}

\newcommand{\tr}{\ensuremath{\mathrm{tr}}}

\newcommand{\rd}{\ensuremath{\mathrm{d}}}

\newcommand{\bi}{\ensuremath{\boldsymbol{i}}}
\newcommand{\bj}{\ensuremath{\boldsymbol{j}}}
\newcommand{\bk}{\ensuremath{\boldsymbol{k}}}
\newcommand{\bn}{\ensuremath{\boldsymbol{n}}}

\newcommand{\bx}{\ensuremath{\boldsymbol{x}}}

\newcommand{\bu}{\ensuremath{\boldsymbol{u}}}

\newcommand{\ba}{\ensuremath{\boldsymbol{a}}}
\newcommand{\bb}{\ensuremath{\boldsymbol{b}}}
\newcommand{\bc}{\ensuremath{\boldsymbol{c}}}
\newcommand{\bd}{\ensuremath{\boldsymbol{d}}}
\newcommand{\be}{\ensuremath{\boldsymbol{e}}}
\newcommand{\bg}{\ensuremath{\boldsymbol{g}}}

\newcommand{\br}{\ensuremath{\boldsymbol{r}}}

\newcommand{\bA}{\ensuremath{\mathbf{A}}}
\newcommand{\bB}{\ensuremath{\mathbf{B}}}

\newcommand{\calA}{\ensuremath{\mathcal{A}}}

\newcommand{\calL}{\ensuremath{\mathcal{L}}}
\newcommand{\calV}{\ensuremath{\mathcal{V}}}

\newcommand{\bI}{\ensuremath{\mathbf{I}}}
\newcommand{\bJ}{\ensuremath{\mathbf{J}}}
\newcommand{\bK}{\ensuremath{\mathbf{K}}}
\newcommand{\bL}{\ensuremath{\mathbf{L}}}

\newcommand{\bM}{\ensuremath{\mathbf{M}}}

\newcommand{\bP}{\ensuremath{\mathbf{P}}}

\newcommand{\bT}{\ensuremath{\mathbf{T}}}

\newcommand{\grad}{\ensuremath{{\mathrm{grad}}}}

\newcommand{\mskew}{\ensuremath{\mathrm{skew}}}

\usepackage{newtxtext}
\usepackage{newtxmath}

\usepackage{tensor}
\usepackage{orcidlink}

\begin{document}
\title{Integral theorems for the gradient of a vector field, with a fluid dynamical application}

\author{Jonathan M.  Lilly\,\orcidlink{0000-0001-5651-7496}$^{1}$, Joel Feske\,\orcidlink{0000-0001-5617-1405}$^{2}$, Baylor Fox-Kemper\,\orcidlink{0000-0002-2871-2048}$^{2}$, and Jeffrey J. Early\,\orcidlink{0000-0003-4332-4569}$^3$}

\address{$^{1}$Planetary Science Institute, Tucson, AZ 85719, USA\\$^{2}$Department of Earth, Environmental, and Planetary Sciences, Brown University, Providence, RI 02912, USA\\$^{3}$NorthWest Research Associates, Seattle, WA 98105 USA}
\subject{}
\keywords{}
\date{today}
\corres{Jonathan M. Lilly\\\email{jmlilly@psi.edu}}

\begin{abstract}
The familiar divergence and Kelvin--Stokes theorem are generalized by a tensor-valued identity that relates the volume integral of the gradient of a vector field to the integral over the bounding surface of the outer product of the vector field with the exterior normal.  The importance of this long-established yet little-known result  is discussed.  In flat two-dimensional space, it reduces to a relationship between an integral over an area and  that over its bounding curve, combining the 2D divergence and Kelvin--Stokes theorems together with two related theorems involving the strain, as is shown through a decomposition using a suitable tensor basis.  A fluid dynamical application to oceanic observations along the trajectory of a moving platform is given.   The potential generalization of the generalized identity to curved two-dimensional surfaces is considered and is shown not to hold.  Finally, the paper includes a substantial background section on tensor analysis, and presents results  in both symbolic  notation and index  notation in order to emphasize the correspondence between these two notational systems.  
\end{abstract}

\begin{fmtext}
\end{fmtext}
\maketitle

\section{Introduction}

Nearly a century and a half ago, the mathematical framework of vector analysis was laid out by Gibbs \cite{gibbs} in a form that has changed little to this day. In that treatise, in a list including well-known integral theorems such as the divergence theorem and Kelvin--Stokes theorem, Gibbs also presents a theorem linking the volume integral  in three-dimensional space of the gradient of a vector field to an integral over the bounding surface, see eqn. (2) in \S161 of \cite{gibbs}.  That result, which we will refer to as the \emph{gradient tensor theorem}, is given by
\begin{equation}
\iiint_\calV  \bm\nabla \otimes \bu\, \rd V=\iint_\calA  \bn \otimes \bu\, \rd A
 \label{3dtensortheorem}
\end{equation} 
for a vector field $\bu$ integrated over a volume $\calV$ with bounding surface $\calA$, and with $\bn$ being the exterior normal vector to $\calA$.  Here the  notation $\ba\otimes\bb$ denotes the outer product of  vector $\ba$ with  vector $\bb$, yielding a tensor, while the tensor-valued gradient of the vector field $\bu$ is denoted as $\bm\nabla \otimes \bu$ and not $\bm\nabla \bu$ for reasons to be discussed subsequently.  This result involves a $3\times 3 $ tensor on each side of the equality, thus comprising a set of nine individual identities, with each component of $\bu$ differentiated with respect to each of the three coordinates.  It contains within itself more familiar identities: its trace is the divergence theorem---obtained by replacing the  outer product  ``$\otimes$'' with the  inner or dot product ``$\cdot$''---while its skew-symmetric part is a vector-valued analogue of the Kelvin--Stokes theorem, together accounting for four of the nine terms.  The remaining identities involve what would be called the rates of strain if $\bu$ is interpreted as a velocity; these too are subject to integral theorems relating their volume integrals to integrals over the boundary.  

The two-dimensional version of the gradient tensor theorem is of particular importance in  fluid dynamics, where two-dimensional flows are a commonplace idealization.  If the vector field $\bu$ lies entirely within a horizontal surface,   (\ref{3dtensortheorem}) reduces to
\begin{equation}
\iint_\calA  \bm\nabla \otimes \bu\, \rd A=\oint_\calL  \rd \bn \otimes \bu \label{flowtheorem}
\end{equation}
which describes the connections between spatially integrated derivatives of the vector field $\bu$ over the area $\calA$ and its value  along the bounding curve $\calL$.  Here  the boundary integral is traversed in the right-hand sense and $\rd\bn$ is the differential exterior normal, related to the differential  tangent vector $\rd \bx$  to the boundary $\calL$ via $\rd\bn=-\bk\times \rd\bx$ where $\bk$ is the vertical unit vector.  For the case of a Cartesian coordinate system, with the two-dimensional vectors $\bx$ and $\bu$ having components $(x,y)$ and $(u,v)$ respectively, one obtains
\begin{equation}
\begin{array}{ll}\displaystyle  \iint_\calA \displaystyle\delta\, \rd A = \oint_\calL\left(-v \rd x+u \rd y  \right) &\quad\quad \displaystyle
\iint_\calA \zeta\, \rd A  = \oint_\calL\left(u \rd x + v \rd y\right)       \vspace{.1in}\\
 \displaystyle \iint_\calA \nu\, \rd A =\oint_\calL \left(v \rd x+u \rd y \right)&\quad\quad\displaystyle\iint_\calA\sigma\,\rd A =\oint_\calL\left(-u \rd x + v \rd y\right)
\end{array}\label{flowexpansionscalar}
\end{equation}
where  $\delta$, $\zeta$, $\nu$, and $\sigma$ are the divergence, the component of vorticity normal to the surface, and the normal strain rate and shear strain rate within the surface, respectively, which may be defined as 
\begin{equation}
\delta \equiv \frac{\partial u}{\partial x} +\frac{\partial v}{\partial y},\quad\quad
\zeta \equiv \frac{\partial v}{\partial x} -\frac{\partial u}{\partial y},\quad\quad
\nu\equiv \frac{\partial u}{\partial x} -\frac{\partial v}{\partial y},\quad\quad
\sigma \equiv\frac{\partial v}{\partial x}+ \frac{\partial u}{\partial y}.  \label{dzns}
\end{equation}
Note that (\ref{flowexpansionscalar})  are all manifestation of Green's theorem, $\iint_\calA \left(\frac{\partial M}{\partial x}-\frac{\partial L}{\partial y}\right)\rd x\rd y = \oint_\calL \left(L \,\rd x +M \,\rd y\right)$.  Thus the two-dimensional gradient tensor theorem combines into one compact and coordinate-independent identity the divergence theorem, the Kelvin--Stokes theorem, and two related identities involving the strain rates.  

The 3D and 2D gradient tensor theorems, (\ref{3dtensortheorem}) or (\ref{flowtheorem}), make no reference to a coordinate system, and are true regardless of which basis is used to represent vectors and tensors.  Nevertheless, is it instructive to write them out in component form for Cartesian coordinates, in which case they will be shown to become
\refstepcounter{equation}
\begin{equation*}
\iiint_\calV \frac{\partial}{\partial x^i} u^j\, \rd V=\iint_\calA n^i  u^j \, \rd A,\quad\quad\quad\quad\quad\quad
\iint_\calA  \frac{\partial}{\partial x^i}u^j\, \rd A=\int_\calL  \rd n^{i} \, u^j  
 \eqno{(\theequation{\mathit{a},\mathit{b}})} \label{cartesianforms}
\end{equation*} 
with $i$ and $j$ each taking on the values $(1,2,3)$ in the former equation and $(1,2)$ in the latter.  Here $x^i$ is the $i$th Cartesian coordinate while $u^i$  denotes the $i$th component of the vector $\bu$.   These forms show the connection to the fundamental theorem of calculus more clearly.  For  a volume consisting of a rectangular cuboid aligned such that its six faces are each normal to one of the basis vectors, the $n^i$ terms in (\ref{cartesianforms}a) become unity or zero on each of the faces, and the equality follows from the fundamental theorem of calculus.  Informally, we can extend this identity to any volume by considering it to be composed of a sum of such cuboids, or see \S 3.31 of \cite{aris} for an approach involving projections onto coordinate planes.  As we are working in three-dimensional space, for (\ref{3dtensortheorem}), or on any flat two-dimensional surface embedded in three-dimensional space, for  (\ref{flowtheorem}), the Cartesian coordinates and the attendant orthogonal basis vectors are always available, and therefore we could always choose to write (\ref{cartesianforms}).  And because we can represent the same vector or tensor objects equivalently in different coordinate systems, we infer the underlying results must be independent of the choice of coordinate system, and this is what  (\ref{3dtensortheorem}) and (\ref{flowtheorem}) express.

The divergence and Kelvin--Stokes theorem are central to the study of fluid dynamics, electrodynamics, and other branches of physics.   They are frequently used in the manipulations of dynamical equations---giving rise to well-known results such as Kelvin's circulation theorem, Gauss's law, and Faraday's law---and can also be usefully applied in observational or model analysis in order to infer spatial averages from information along a boundary.  But while these integral theorems are common knowledge, this does not appear to be the case for the gradient tensor theorem.  In the roughly two dozen textbooks on the first author's bookshelf on  vector and tensor analysis, classical dynamics, electrodynamics, fluid dynamics, and physical oceanography, including classic works such as \cite{aris,panton,kundu,badin,borisenko,truesdell,batchelor}, the gradient tensor theorem in one of the coordinate-independent forms (\ref{3dtensortheorem}) or (\ref{flowtheorem}) could not be located.   One may however find the result $\iiint_\calV  \frac{\partial}{\partial x^i} \varphi \, \rd V=\iint_\calA n_i  \varphi \, \rd A$,  where $\varphi$ may be a scalar or a component of a vector or tensor, see (3.12.2) of Panton \cite{panton}, (2.30) of Kundu et al. \cite{kundu}, and (3.31.2) of Aris \cite{aris}\footnote{The authors are grateful to an anonymous reviewer for bringing these equations to our attention.}, which indeed becomes the Cartesian componentwise identity (\ref{cartesianforms}a) with the choice $\varphi=u^j$. Furthermore, although they are certainly not new, we have not been able to find in the literature the two strain theorems that augment the more familiar divergence and Kelvin–Stokes theorems in the two-dimensional gradient tensor theorem (\ref{flowexpansionscalar}), despite the fact that they immediately follow from (\ref{cartesianforms}b).   These factors suggest that the gradient tensor theorem could be more widely known and appreciated than it is at present.  This theorem does however appear to be standard in continuum mechanics, where it may be found in several recent textbooks [\citen{ogden}, eqn. (1.5.67); \citen{dimitrienko}, eqn. (3.23); \citen{murakami}, eqn. (12.170)].

The purpose of this work is to reintroduce the gradient tensor theorem to the larger community of physical scientists, to explain its importance, and to present an application.  In so doing, it is necessary to also treat the outer product, as an understanding of this operator is essential in order for (\ref{3dtensortheorem}) and (\ref{flowtheorem}) to be meaningful.  In fluid mechanics the outer product, like gradient tensor theorem itself, does not appear to be a part of the standard lexicon.   This is unfortunate because the tensor-valued gradient of a vector field arises frequently, and is implicit within the ubiquitous nonlinear advection term $(\bu\cdot\bm\nabla)\bu$.  Often in textbooks the symbol $\bm\nabla \bu$, rather than being defined, is left to be interpreted in the context of expressions such as $(\bu\cdot\bm\nabla)\bu$ [e.g., \citen{batchelor,mcwilliams,saffman,battaner}]. Yet without recognizing that the gradient   $\bm\nabla\bu$ is standing in for the outer product derivative $\bm\nabla\otimes\bu$, we cannot remove the parenthesis from the advection term to write $(\bu\cdot\bm\nabla)\bu=\bu\bm\nabla\otimes\bu$. As will be seen subsequently, this identity is in fact the defining property of the ``$\otimes$'' operator.   Understanding the gradient as the outer product  $\bm\nabla\otimes\bu$ therefore goes hand in hand with understanding the gradient tensor theorem.  Moreover, the appreciation of  $\bm\nabla\bu$ as being an implicit outer product also resolves a transposition ambiguity in definition of the gradient tensor that exists in the literature, as pointed out recently by \cite{wood22-arxiv}.

Several questions then arise regarding the interpretation and range of validity of the gradient tensor theorem.  It is natural to ask  whether the two-dimensional gradient tensor theorem (\ref{flowtheorem}) is also valid for curved surfaces embedded in three-dimensional space, such as the surfaces of constant density encountered frequently in fluid mechanics.  It will be shown that the answer is no, it only holds for flat surfaces, although the divergence theorem and Kelvin--Stokes theorem do still apply.  Similarly, one may ask whether the gradient tensor theorem is contained within, or implied by, the generalized Stokes theorem of differential geometry---a result that embodies the essence of Stokes-like integral theorems within a broad generalization of what is meant by a volume and a boundary.  Again the answer will be shown to be no: while the generalized Stokes theorem and the gradient tensor theorem can both be seen as generalizations of the classical Kelvin--Stokes theorem, and are superficially similar, they represent different generalizations.  

A work presenting a result involving tensors must confront the reality that familiarity with tensor analysis is not as widespread as it deserves to be, particularly in comparison with vector analysis.  Whereas vector analysis appears in introductory texts on physics, electrodynamics, and fluid mechanics, tensors are typically reserved for more advanced topics such as classical mechanics and especially for general relativity.  As a consequence, one is faced with a choice: present the result tersely, thus limiting the readership to those already familiar with tensors, or take the time to present the background material required by a larger audience.  Here we take the latter approach.   This is particularly suitable for this topic, because as will be seen, the story of the gradient tensor theorem is one that is interwoven with the history of tensor analysis.  

The choice then arises of what notation system to use for tensor analysis.  One option is to use the index notation system, as is standard in general relativity \cite{carroll} and advanced physics \cite{grinfeld}. This is the system formulated by Ricci-Curbastro  and Levi-Civita \cite{ricci00-ma,levi-civita} in their foundational work on modern tensor analysis, streamlined by the summation convention introduced by Einstein \cite{einstein16-adp}.  A second option is to use a symbolic system derived from that of Gibbs \cite{gibbs}, which involves the use of the familiar $\bm\nabla$ operator. Updated and modernized versions of this system, including the incorporation of curvilinear coordinates,  are found in recent continuum mechanics  texts such as\cite{carroll,itskov,dimitrienko,murakami}.   The former system has the advantages of compactness and explicitness that allow manipulations to be carried out with ease.  The latter has the virtue of familiarity, as it is essentially continuous with the ubiquitous notation for vector analysis while mirroring the mechanics of linear algebra, as well as a perhaps more ready legibility.   Reviewing the situation for fluid mechanics, Panton \cite[p. 29]{panton} concludes ``Workers in ﬂuid mechanics must have a knowledge of both symbolic and index notations. The two notations are used with equal frequency in the literature.''  As both systems are in common use,  we present the main results in parallel using both notations.   At the same time, in order to ensure accessibility, the sections that rely on index notation, in particular \S\ref{prelim}\ref{curvilinear} and   \S\ref{derivation}\ref{curveresults}, are marked with an asterisk; the remainder of the paper may be read without these two sections. 

This brings us to a secondary goal of the paper.  In seeking to determine the implications and possible extensions of the gradient tensor theorem, we have found a solid understanding of both symbolic and index notation to be essential.  The process of carrying out this work has emphasized to us their distinct strengths and weaknesses, highlighted areas in which these two ultimately equivalent notations appear surprisingly distant, and underscored the value of being conversant in both.   The background section could therefore be read on its own as a very compact introduction to tensor analysis that treats symbolic and index notation even-handedly and that emphasizes their deep congruency, something that is rare in the literature.  It is our hope that this presentation will benefit others working in tensor analysis, and also help to make this powerful tool more broadly accessible.

This paper  builds on recent work by the first author \cite{lilly18-fluids}, in which a version of the two-dimensional gradient tensor theorem, (\ref{flowtheorem}), was derived in the framework of linear algebra, by representing vectors as $2\times 1$ arrays and tensors as  $2\times2$ matrices.  Here we revisit the problem within the context of the more abstract and general framework of tensor analysis, with an appreciation of how the tensor gradient theorem fits into a hierarchy of related identities, and with a perspective on the origin and potential usefulness of this result.

The structure of the paper is as follows.  As motivation, an application to a real-world problem in observational oceanography is presented first in \S\ref{application}. Preliminaries centering on vector and tensor analysis involving the outer product  are then given in \S\ref{prelim}.   A derivation of the two-dimensional gradient tensor theorem starting with the divergence theorem is presented in \S\ref{derivation}, followed by  an analysis of its components.  Key contributions are the identification of a tensor basis that lets the two-dimensional gradient tensor theorem (\ref{flowtheorem}) be readily decomposed into the four physically meaningful constituent identities given in (\ref{flowexpansionscalar}), thus unlocking the practical value of this result, and a demonstration that the two-dimensional gradient tensor theorem does not apply to curved surfaces; these are presented in \S\ref{derivation}\ref{twodgradtensor} and  \S\ref{derivation}\ref{curveresults} respectively.   The paper concludes  in \S\ref{discussion} with  a discussion. 

\section{An application}\label{application}

The gradient tensor theorem is likely to be of particular use in fluid mechanics, where the vector field $\bu$ being studied is generally the velocity and where the velocity gradient tensor is an object of great interest.   It is known that the spatial derivatives the velocity field play essential roles in controlling the local evolution of the flow \cite{okubo70-dsr,kirwan75-jpo,cantwell92-pfa,meneveau11-arfm,samelson13-arms}, as numerical or laboratory modelling diagnostics of local topology and dissipation \cite{cantwell93-pfa,martin98-pf,bechlars17-jfm,buxton17-jfm,wu20-pf,atkinson21-pf}, and as controlling terms within energy balance equations  \cite{zemskova15-jpo,whitt15-jpo,lelong20-jpo}.  Because of this importance, in the study of the ocean currents considerable effort has been expended in the estimation of the velocity gradient terms from in situ measurements \cite{molinari75-jpo,okubo76-dsr,flament00-jpo,rudnick01-grl,shcherbina13-grl,ohlmann17-grl,berta20-jgr,oscroft20-fluids,tarry22-grl,essink22-jaot}. In the analysis of the velocity gradient tensor from observations or within numerical models, the ability to infer spatial averages from information along boundaries may prove to be invaluable, for example, by providing ready access to aggregate statistics or possibly by facilitating the formulation of integral evolution laws.   This is particularly evident when one considers the fact  that key high-resolution observational platforms in  oceanography,  such as ship-based surveys \cite{rudnick01-grl,shcherbina13-grl}, the now widely-used autonomous gliders  \cite{hatun07-jpo,rudnick15-jpo,yu17-jgr,meunier18-jgr}, and the new Saildrone  \cite{zhang19-oceanography} and SailBuoy \cite{wullenweber22-sensors} autonomous surface platforms, all inherently sample along a trajectory and not over an area. 

\begin{figure}
\begin{center}
\includegraphics[width=\textwidth,angle=0]{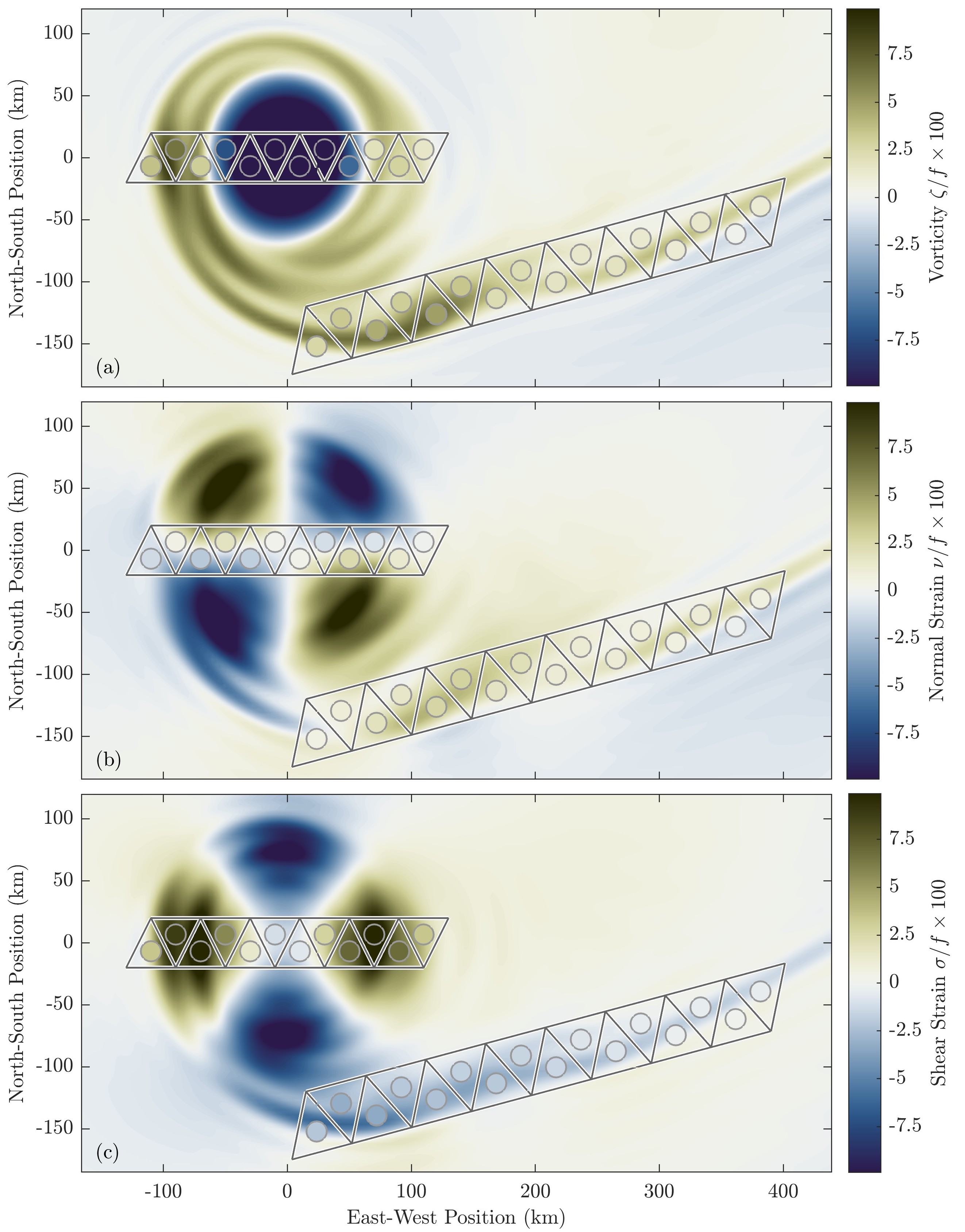}
\end{center}
	\caption{An illustration of the use of the gradient tensor theorem in observational oceanography.  The  (a) vorticity $\zeta$, (b) normal strain $\nu$, and (c) shear strain $\sigma$ in a snapshot of a numerical simulation of a quasigeostrophic eddy termed BetaEddyOne \cite{early23-zenodo} are shown, each normalized by the Coriolis frequency $f\equiv 2\Omega \sin \phi$ where $\Omega$ is the angular rotation rate of the Earth and $\phi$ is the model's central latitude.  The eddy is sampled along the triangular tracks, as described in the text, and the colours of the circular disks show the average values of the three quantities within each triangular cell as inferred from the information along its boundary using (\ref{flowtheorem}).  Note that the divergence vanishes for this quasigeostrophic model.  See the Data Accessibility section at the end of the paper for details on model output availability. }\label{vortex}
\end{figure}

To illustrate the relevance to ocean observations, an application to a long-lived eddy in a numerical simulation termed BetaEddyOne \cite{early23-zenodo} is shown in figure~\ref{vortex}, sampled in such a way as to mimic a one-dimensional observational platform.   Inspired by the study of \cite{early11-jpo}, an initial condition of a circular eddy is placed in quiescent ocean and integrated under 1.5 layer quasigeostrophic dynamics for one year.  Eddy features such as this one are known to be important players in the climate system \cite{chelton11-pio,abernathey18-jpo,zhang20-fluids} but are difficult to observe remotely due to their small size relative to the resolution of satellite-based platforms.  In situ measurements \cite{hatun07-jpo,yu17-jgr,meunier18-jgr} are therefore essential for understanding the details of these eddies. 

Assuming the model eddy remains frozen during the time it takes the observations to be carried out, the eddy core and a trailing filament are each observed along a track composed of closed triangular cells.  Such tracks can be drawn by a single moving platform, without overwriting any lines, by beginning at an upper vertex and then moving southwest, due east, northwest, and due east in succession.  While only two terms from the velocity gradient tensor are known instantaneously along a moving instrument trajectory, average values of the entire tensor are known within each triangular cell from the information along the boundary using the two-dimensional gradient tensor theorem  (\ref{flowtheorem}). The vorticity $\zeta$, normal strain $\nu$, and  shear strain $\sigma$ inferred in this way are seen to provide good approximations to the full local structure; note that the divergence vanishes for the quasigeostrophic dynamics employed in this simulation.  This demonstrates that accurate small-scale velocity gradient information can be obtained with measurements from a single ship or glider using the gradient tensor theorem, a fact which could improve our ability to observe such structures. 

\begin{figure}
\begin{center}
\includegraphics[width=\textwidth,angle=0]{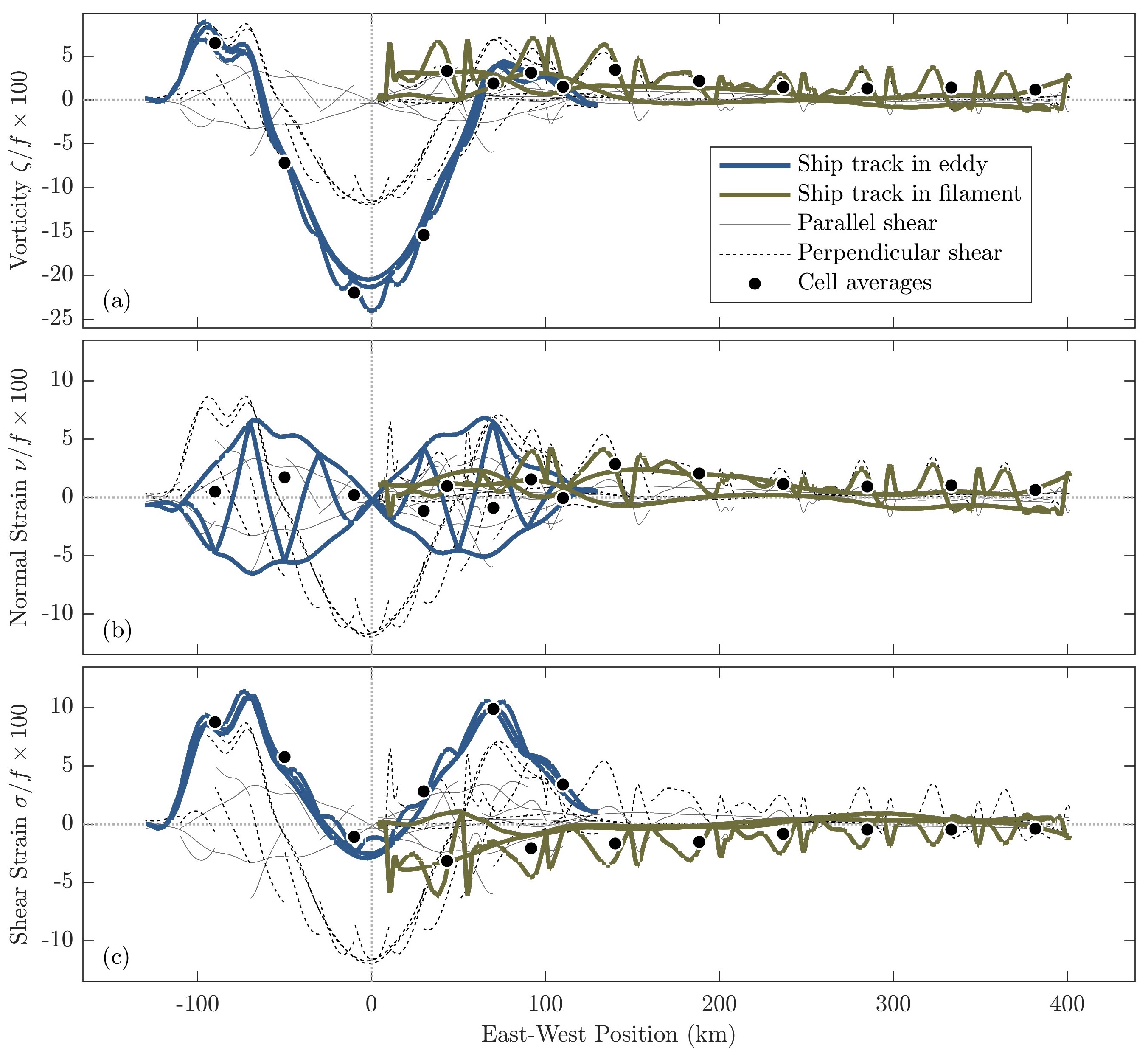}
\end{center}
	\caption{ Another view of the information presented in figure~\ref{vortex}.  Here the instantaneous values of the  (a) vorticity $\zeta$, (b) normal strain $\nu$, and (c) shear strain $\sigma$ along the sampling tracks are shown, as found from interpolating within the model.  These are not directly observable because the gradient can only be computed in the along-track direction.  For reference, velocity gradient components locally parallel and perpendicular to the sampling tracks are shown in each panel, indicating the information that can be recovered directly from taking along-track derivatives.  Dots show the cell-averaged values, generally providing a good match to the local structure, apart from in panel (b) where the normal strain $\nu$  exhibits small-scale oscillatory behaviour.  }\label{profiles}
\end{figure}

Another view of this application is found in the line plot in figure~\ref{profiles}.  Along the moving platform, the velocity gradient terms $\frac{\partial u}{\partial \tilde x}$ and $\frac{\partial v}{\partial \tilde x}$ can be formed, where $\tilde x$ is the position coordinate along the track, but  spatial derivatives in the direction perpendicular to the track are not known.  The two observable terms can be rotated to give $\frac{\partial \tilde u}{\partial \tilde x}$ and $\frac{\partial \tilde v}{\partial \tilde x}$, the along-track derivatives of the velocity components parallel and perpendicular to the track, respectively.  These are shown in  figure~\ref{profiles} and compared with the full gradient terms  $\zeta$, $\nu$, and  $\sigma$ along the tracks.  This illustrates that having the along-track information is insufficient to reconstruct $\zeta$, $\nu$, and  $\sigma$, as each of these is a sum or difference of two terms, only one of which is known.  In  this particular case, we can exploit the fact that the divergence vanishes to recover the normal strain in the reference frame aligned with the track direction as  $\tilde\nu=2\frac{\partial \tilde u}{\partial \tilde x}$, but we cannot estimate the $\frac{\partial \tilde u}{\partial \tilde y}$ term that appears in both the vorticity and shear strain without resorting to geometric assumptions exploiting the near-circular shape of the eddy.  The triangle-cell averages, however, give generally representative values of the instantaneous values of the gradient terms, as well as recovering their areal averages.  

For  the particular case of the normal strain $\nu$ in  figure~\ref{vortex}b, values along the instrument track oscillate between  positive and negative, leading to cancellations within the triangular cells; the cell averages in figure~\ref{profiles}b, while accurate, are thus less representative of this rapidly varying along-track structure.  This illustrates the caveat that interpreting results from the gradient tensor theorem is more straightforward when some degree of local smoothness may be assumed.  More generally, in the application of integral theorems one must keep in mind that if the field one is sampling has variability on the scale of the sampling, this could lead to cancellations wherein the integral value underestimates the local value.

\section{Background}\label{prelim}

In this section we present necessary concepts and notation from vector and tensor analysis, in particular the outer product and associated derivatives, limiting our consideration to ordinary three-dimensional space.   As described in the Introduction, we will work with both symbolic notation and index notation.   The symbolic perspective will be presented first, initially limited to Cartesian coordinates, in which we emphasize (i) the distinction between proper vectors and tensors and the arrays of numbers used to represent them, as stressed by Koks~\cite{koks17-dstg}, and (ii) the parallel between tensors operations and those in linear algebra, as in the presentation of Itskov~\cite{itskov}.  This is followed by a treatment of more general curvilinear coordinates using both index notation and symbolic notation. See  \cite{koks17-dstg,borisenko,ogden,murakami,itskov} for further details.  To facilitate broad accessibility, concepts from differential geometry, found integrated with tensor analysis in works such as \cite{carroll,needham}, are intentionally avoided.  As this section is intended as background, those readers already familiar with this material are invited to turn immediately to the results section.   

\subsection{Vector and tensor notation}\label{tensorappendix}

Recall that a vector, in physics, is an abstract entity having both a magnitude and direction that is independent of any frame of reference or choice of coordinates.  Two vectors  $\ba$  and $\bb$  can be multiplied to give a scalar via the inner product, defined as $\ba\cdot\bb \equiv \|\ba\|\|\bb\| \cos \vartheta$ where $\|\ba\|$ and $\|\bb\|$ are the lengths of the two vectors and $\vartheta$ is the angle between them.  A second-order tensor $\bM$ can be defined as a linear mapping that, given any vector $\ba$, yields a new vector $\bb$, a correspondence that is written as $\bb=\bM\ba$. By linear we mean that $\bM(\ba+\bb)=\bM\ba+\bM\bb$ together with $\bM(\alpha\ba)=\alpha\left(\bM\ba\right)$ for any scalar $\alpha$. Second-order tensors will be referred to simply as tensors herein.  Like vectors, tensors, as linear mappings of one vector to another, are independent of our choices of frames of reference and coordinate systems.   

To achieve concrete representations of vectors and tensors as arrays of numbers, we must first choose some coordinate system within a particular frame of reference.   Given a Cartesian coordinate system $S$ with coordinates $x^i$ and associated basis vectors $\be_i$ for $i=1,2,3$, we can expand the vectors $\ba$ and $\bb$ as $\ba=a^1\be_1+a^2\be_2+a^3\be_3$ and $\bb=b^1\be_1+b^2\be_2+b^3\be_3$, where vector components are labelled by superscripts rather than subscripts for reasons to become apparent shortly. These coefficients can be gathered into column vectors, while as shown subsequently  the tensor $\bM$ can be expressed as a matrix, leading to
\begin{equation}
\left[\ba\right]_S = \begin{bmatrix} a^1 \\ a^2\\ a^3\end{bmatrix},\quad\quad\quad
\left[\bb\right]_S = \begin{bmatrix} b^1 \\ b^2\\ b^3\end{bmatrix}, \quad\quad\quad
\left[\bM\right]_S  =  \begin{bmatrix} M^{11} & M^{12} & M^{13}\\  M^{21} & M^{22} & M^{23}\\  M^{31} & M^{32} & M^{33}\end{bmatrix}.\label{representations}
\end{equation}  
Here the notation $\left[\cdot\right]_S$,  following e.g. \cite{koks17-dstg} and \S 2.7 of \cite{bradley}, denotes a representation with respect to coordinate system $S$. It is worth emphasizing that if we change coordinate systems, $\ba$, $\bb$, and $\bM$ are unaffected while $[\ba]_S$, $[\bb]_S$, and $[\bM]_S$ certainly do change. To distinguish quantities like $\ba$ from those like $[\ba]_S$, both of which go by the name ``vector'', we refer to quantities of the former type as \emph{proper vectors} or simply \emph{vectors}, while those of the latter will be termed  \emph{representation vectors}.   The inner product $\ba\cdot\bb$ and the tensor operation $\bM\ba$ then have representations
\begin{equation}\label{tworepresentations}
\ba\cdot \bb = [\ba]_S^T[\bb]_S,\quad\quad\quad [\bM\ba]_S = [\bM]_S[\ba]_S
\end{equation}  
with the superscript ``$T$'' denoting the transpose.  The former follows from  the assumed orthonormality of the basis vectors in a Cartesian coordinate system, $\be_i\cdot \be_j=\delta_{ij}$ where $\delta_{ij}$ is the Kronecker delta function, and the latter from the stipulation that $\bM\ba$ is a linear mapping of vectors to vectors. 

The above presentation, which closely follows that of Itskov \cite{itskov}, has led to a notational system which mirrors that of familiar linear algebra---vectors $\ba$ and $\bb$ and tensors $\bM$ act as if they were column vectors and matrices, respectively, despite being abstract entities  independent of any representation.  To continue this development, we may also  define the operation of $\bM$ on vectors to its left by demanding that 
\begin{equation}\label{leftoperation}
\bb\cdot(\bM\ba)=(\bb \bM)\cdot \ba
\end{equation}
and with this definition of left operation, we may dispense with the dots and parentheses and write this product unambiguously as $\bb\bM\ba$.  The transposed tensor, denoted $\bM^T$,  is similarly defined such that 
\begin{equation}\label{tensortranspose}
(\bM \ba)\cdot \bb = \left(\bM^T \bb\right)\cdot \ba.
\end{equation}  
In other words, if one operates on $\ba$ with a tensor $\bM$ and then projects the results onto $\bb$, yielding a scalar, $\bM^T$ is defined as the mapping that will give the same answer if one instead operates first on $\bb$ and then projects the result onto $\ba$.  Note that the tensor transpose has been defined abstractly as an operation without any explicit notion of changing shape.   From this definition it follows that if $\bM=\bA\bB$, then $\bM^T=\bB^T\bA^T$.    The representation for the product  $\bb\bM\ba$ is then $\bb\bM\ba = [\bb]_S^T[\bM]_S[\ba]_S$ while that of the transposed tensor is $[\bM^T]_S=[\bM]_S^T$, so that these tensor operations again behave just like the familiar linear algebra operations on their representations.  Unlike the linear algebra case, however, no meaning is ascribed to the transposition of proper vectors; whereas $[\ba]_S^T$ indicates changing from a column vector to a row vector, $\ba^T$ is not defined. 
 
At this point we introduce a notational change. We adopt the Einstein summation convention that a summation over an index is implied whenever the same index appears as a pair in both an upper and lower position.   This allows us to write the vector $\ba$ compactly as $\ba=a^i\be_i$ rather than $\ba=\sum_{i=1}^3a^i\be_i$.  Moreover, for notational convenience in Cartesian coordinates we define $\be^i\equiv\be_i$ for the basis vectors, $a_i\equiv a^i$ for vector components, and $\delta^{ij}=\delta_i^j=\delta^i_j=\delta_{ij}$ for the Kronecker delta.  We can then write an inner product $\ba\cdot \bb$ as
 \begin{equation}
\ba\cdot\bb= a^i \be_i \cdot b_j \be^j = a^ib_j \delta_i^j=a^ib_i
 \end{equation}
without explicit summations. The use of upper and lower indices with the Cartesian basis is purely notational as $\be^i$ and $\be_i$ are momentarily simply different names for the same quantity, and similarly for $a^i$ and $a_i$. We emphasize that index position has nothing to do with transposition, which is not defined for proper vectors; conversely, transposition of representation vectors does not affect index position.  This convention has the advantage of creating continuity with the index notation used for curvilinear coordinates, wherein the distinction between upper and lower indices takes on an important meaning.

\subsection{The outer product}\label{tensorappendix}

Central to understanding the gradient tensor theorem is the outer product operation, which we now review.  Unlike the case of two scalars, which can be multiplied together in only one way, there are multiple ways that two vectors $\ba$ and $\bb$ can be multiplied together.  The notations
\begin{equation}
\ba\cdot \bb,\quad\quad\quad \ba\times \bb,\quad\quad\quad \ba\otimes \bb \label{threemults}
\end{equation}
denote the inner or dot product, the cross product, and the outer product, respectively, corresponding in turn  to a scalar, a vector\footnote{The quantity formed from a cross-product  $\ba\times \bb$ is more precisely known as an axial vector or pseudovector, as distinguished from a polar vector or true vector. When the universe is rotated, both polar and axial vectors will rotate in exactly the same way; but if the universe is reflected, as in a mirror, then axial vectors will undergo an additional sign change.  This difference in behavior can be seen as arising from the fact that the definition of a cross product involves a sign choice that is a matter of convention, settled by the so-called right-hand rule; but in a mirror, right hands are transformed into left hands. Herein the term \emph{vector} is used to refer to both polar and axial vectors.}, and a tensor.  While the former two are standard, the latter is perhaps less familiar.  The outer product of two vectors $\ba\otimes \bb$ is defined in terms of its action on a third vector $\bc$ such that\begin{equation}
\left(\ba\otimes \bb \right) \bc =\ba \left(\bb \cdot \bc \right)\label{oprodform}
\end{equation}
meaning that  $\ba\otimes \bb$ maps $\bc$ into a new vector that will be parallel to $\ba$ with a coefficient given by the projection of $\bc$ onto $\bb$.    Thus  $\ba\otimes \bb$ is a linear mapping of one vector onto another vector, that is, a tensor.   

It follows from this definition of the outer product  that the array representation of an outer product in a Cartesian coordinate system $S$ must be $\left[\ba\otimes\bb\right]_S=[\ba]_S[\bb]_S^T$, because then we have
\begin{equation}
\left[\left( \ba\otimes\bb\right)\bc\right]_S =\left[ \ba\otimes\bb\right]_S\left[\bc\right]_S = [\ba]_S[\bb]_S^T[\bc]_S = [\ba (\bb \cdot \bc)]_S
\end{equation}
as desired.  This  leads to the following  comparison between the representations of inner and outer products:
\refstepcounter{equation}
\begin{equation*}
 \ba\cdot\bb = [\ba]_S^T[\bb]_S =a^ib_i,\quad\quad\quad
\left[ \ba\otimes\bb\right]_S = [\ba]_S[\bb]_S^T = \begin{bmatrix} a^1b^1 & a^1 b^2 & a^1 b^3\\
 a^2b^1 & a^2 b^2 & a^2 b^3 \\ 
 a^3b^1 & a^3 b^2 & a^3 b^3
  \end{bmatrix}.
 \eqno{(\theequation{\mathit{a},\mathit{b}})}\label{innerandouter}
\end{equation*}
Note that the former corresponds to a single number while the latter corresponds to a $3\times 3$ matrix.  Thus following the rules of linear algebra, the difference between the  array representations of the inner and outer products is whether we transpose the first or the second array in forming their product.

Outer products are significant because they constitute a minimal type of tensors, with the outer products of the basis vectors forming the building blocks of general tensors.  In three dimensions we have
\begin{equation}
\bM = M^{ij} \be_i \otimes \be_j
\label{matrixexpand}
\end{equation}
where the summation convention implies a sum over all values of both $i$ and $j$.  To see this, we observe that if we form the quadratic products $\be^i\bM\be^j$, the proposed form of $\bM$ yields the matrix given in (\ref{representations}) from orthogonality.  Thus any tensor can be represented as a weighted sum of the nine outer products of the basis vectors with one another.   This  together with  $\ba=a^i \be_i$ leads to  $[\bM\ba]_S=[\bM]_S[\ba]_S$ as stated earlier in (\ref{tworepresentations}). 

Several other properties of outer products will be needed.  From the definition of left operation in (\ref{leftoperation}), it must be the case that  the outer product operates to its left as
\begin{equation}
\bc\left(\ba\otimes \bb \right) =\left(\bc \cdot \ba \right)\bb\label{oprodformright}
\end{equation}
since taking the inner product  with a fourth vector $\bd$ then gives $\bc\left(\ba\otimes \bb \right) \cdot\bd=\bc\cdot \left(\ba\otimes \bb \right) \bd=\left(\bc\cdot\ba\right)\left(\bb\cdot \bd\right)$ for the left-hand side, matching what we would obtain for the right-hand side.  From  the definition of the tensor transpose, (\ref{tensortranspose}),  the transpose of an outer product must be
\begin{equation}
\left(\ba\otimes\bb\right)^T=  \bb\otimes\ba\label{dyadtranspose}
\end{equation}
since $\left[(\ba\otimes\bb) \bc \right]\cdot \bd$ and $ \left[(\bb\otimes\ba) \bd \right]\cdot \bc$ both reduce to the same value, namely $(\ba\cdot \bd)(\bb\cdot \bc)$, as desired. Furthermore, two successive outer products collapse to the single outer product,
\begin{equation}\label{twodyads}
\left(\ba \otimes \bb\right) \left(\bc \otimes \bd\right)  = \left(\bb\cdot \bc\right)\left( \ba \otimes \bd\right),
\end{equation}
between the two outermost vectors $\ba$ and $\bd$, scaled by $(\bb\cdot\bc)$.  This \emph{collapse rule} can be verified by operating on another vector $\bx$ to the right and then applying (\ref{oprodform}) to both sides.  

We will  also make use of the tensor \emph{trace}, which can be defined via its action on an outer product as
\begin{equation}
\tr\left\{\ba \otimes \bb\right\}  = \ba \cdot \bb.\label{dyadtrace}
\end{equation}
From this it follows, together with the collapse rule (\ref{twodyads}), that the trace of two successive outer products is
\begin{equation}\label{dyaddyadtrace}
\tr\left\{\left(\ba \otimes \bb\right) \left(\bc \otimes \bd\right)\right\}  = \left(\bb\cdot \bc\right) \left(\ba \cdot \bd\right)
\end{equation}
which together with \eqref{oprodformright} in turn implies 
\begin{equation}\label{dyadtenstrace}
\tr\left\{\left(\ba\otimes\bb\right)\bM\right\}  = \bb\bM\ba 
\end{equation} 
for the trace of an outer product  acting on a general tensor $\bM$,  as shown in the Appendix.   

Finally, we will need some notation for the symmetric and skew-symmetric parts of a tensor. A  tensor $\bM$ is said to be symmetric if $\bM=\bM^T$ and skew-symmetric if $\bM=-\bM^T$, and its skew-symmetric part  is
\begin{equation}
\mskew\!\left\{ \bM \right\}= \frac{1}{2} \left(\bM-\bM^T\right)
\label{skewdef}
\end{equation}
since then $\mskew\!\left\{ \bM \right\}=-\mskew\!\left\{ \bM \right\}^T$ as desired.  The skew-symmetric part of an outer product of two vectors is closely related to their cross product.   Following  \cite{koks17-dstg}, define the \emph{skew tensor} of $\ba$, denoted $\ba^\times$, as the  tensor that operates on another vector $\bb$ to its right to yield the cross product of $\ba$ and $\bb$, 
\begin{equation}
\ba^\times \bb = \ba \times \bb.\label{skewopt}
\end{equation}
The tensor  $\ba^\times$ is necessarily skew-symmetric: we know that $\bc \cdot \ba \times \bb = - \bb \cdot \ba \times \bc$ from the circular shift invariance of the scalar triple product together with the anticommutivity of the cross product, which implies that we must have  $\bc \ba^\times \bb = \bb  \left(\ba^\times\right)^T \bc = -\bb\ba^\times \bc$ by the definition of the tensor transpose.  In the Cartesian coordinate system~$S$, the representation matrix corresponding to the tensor $\ba^\times$ must  be\footnote{For  readers familiar with differential geometry, we point out that  the operator ``$^\times$'' is related to the Hodge star operator  in three dimensions.} 
\begin{equation}
\left[\ba^\times\right]_S= \begin{bmatrix} 0 & -a^3 & a^2\\
 a^3 & 0&  -a^1 \\
-a^2  & a^1 &0 \end{bmatrix}
\label{skewoptexplicit}
\end{equation}
in order to have $\ba^\times \bb = \ba \times \bb$ as desired.   One then finds a correspondence between the skew-symmetric part of an outer product, and the skew tensor of a cross product:
\begin{equation}\label{skewcorrespond}
\mskew\!\left\{ \ba\otimes\bb \right\} =-  \frac{1}{2} \left(\ba \times \bb\right)^\times
\end{equation}
as is  seen by writing out the  representation tensors of both sides in the Cartesian coordinate system~$S$, or by observing that the standard identity $\left(\ba\times \bb\right)\times \bc = \bb\left(\ba\cdot\bc) -\ba(\bb\cdot \bc\right) $ implies $\left(\ba\times \bb\right)^\times =\bb\otimes \ba-\ba\otimes\bb$.  Combining this with $\tr\left\{\ba \otimes \bb\right\}  = \ba \cdot \bb$ from (\ref{dyadtrace}), we see that the outer product $\ba\otimes\bb$ of two vectors embeds within itself both their inner product $\ba\cdot\bb$, as its trace, and their cross product $\ba\times\bb$ within its skew-symmetric part, with the remaining terms constituting its traceless symmetric part.  Thus $\ba\otimes\bb$ may be regarded as the most general way to multiply  two vectors, as pointed out by Gibbs \cite[\S114]{gibbs}.

As an aside, we mention that a tensor consisting of a single outer product $\ba\otimes\bb$ was called a  \emph{dyad} by Gibbs \cite{gibbs}, an archaic term that is nevertheless still occasionally encountered today \cite{vanbladel,murakami,irgens}.  Gibb's notation for an outer product was the juxtaposition $\ba\bb$, but the notation  $\ba\otimes\bb$ today appears to be more standard;  $\ba\bb$ appears now primarily used for the geometric product of geometric algebra \cite{hestenes}, defined as $\ba\bb=\ba\cdot\bb + \ba \wedge \bb$ where $\ba \wedge \bb$ is the wedge product from differential geometry.

\subsection{The gradient tensor}

Turning now to spatial derivatives of a vector field, one encounters a point of potential ambiguity.  There are two different definitions  in use for the tensor-valued gradient of a vector field, denoted herein $\grad_L\!\left\{\bu\right\}$ and $\grad_R\!\left\{\bu\right\}$, the matrix representations of which in a Cartesian coordinate system $S$ are given respectively by 
\refstepcounter{equation}
\begin{equation*}
[\grad_L\!\left\{\bu\right\}]_S=
\begin{bmatrix} 
 \frac{\partial u^1}{\partial x^1} &  \frac{\partial u^2}{\partial x^1} &\frac{\partial u^3}{\partial x^1} \vspace{.03in} \\
 \frac{\partial u^1}{\partial x^2} &  \frac{\partial u^2}{\partial x^2} &\frac{\partial u^3}{\partial x^2} \vspace{.03in} \\
 \frac{\partial u^1}{\partial x^3} &  \frac{\partial u^2}{\partial x^3} &\frac{\partial u^3}{\partial x^3} \end{bmatrix},\quad\quad\quad
 [\grad_R\!\left\{\bu\right\}]_S=
\begin{bmatrix} 
 \frac{\partial u^1}{\partial x^1} &  \frac{\partial u^1}{\partial x^2} &\frac{\partial u^1}{\partial x^3} \vspace{.03in} \\
 \frac{\partial u^2}{\partial x^1} &  \frac{\partial u^2}{\partial x^2} &\frac{\partial u^2}{\partial x^3} \vspace{.03in} \\
 \frac{\partial u^3}{\partial x^1} &  \frac{\partial u^3}{\partial x^2} &\frac{\partial u^3}{\partial x^3} \end{bmatrix}
 \eqno{(\theequation{\mathit{a},\mathit{b}})}\label{graddefs}
\end{equation*}
and which are seen to differ by a transpose, $ [\grad_L\!\left\{\bu\right\}]_S= [\grad_R\!\left\{\bu\right\}]_S^T$.  In the literature one finds both the former [\citen{truesdell}, (3.1);\citen{dimitrienko}, (2.24)] and the latter [\citen{ogden}, (1.5.15); \citen{itskov}, (2.64--2.66); \citen{irgens}, (4.4.17); \citen{meneveau11-arfm}, (2); \citen{buxton17-jfm}, (1.1)] definitions.  These conventions could be seen as arising from whether one defines the gradient operator in terms of the directional derivative, as in e.g. (1.5.3) of \cite{ogden} and (4.4.4) of \cite{irgens}, via a left- or right-handed operation on some other vector $\ba$,
\refstepcounter{equation}
\begin{equation*}
\ba \,\grad_L\!\left\{\bu\!\left(\bx\right)\right\}  = \frac{\rd}{\rd s} \left. \bu(\bx+s\ba)\right|_{s=0},\quad\quad\quad \grad_R\!\left\{\bu\!\left(\bx\right)\right\} \ba  = \frac{\rd}{\rd s} \left. \bu(\bx+s\ba)\right|_{s=0},
 \eqno{(\theequation{\mathit{a},\mathit{b}})}\label{graddefsdirectional}
\end{equation*}
a fact that suggests the terms \emph{left gradient} and \emph{right gradient} for $\grad_L\!\left\{\bu\!\left(\bx\right)\right\}$ and $\grad_R\!\left\{\bu\!\left(\bx\right)\right\} $ respectively.  These in turn imply differences in how one forms the Taylor series expansion,
\refstepcounter{equation}
\begin{equation*}
\bu(\bx)\approx\bu(\bx_o)+\left(\bx-\bx_o\right) \grad_L\!\left\{\bu\!\left(\bx\right)\right\},\quad\quad\quad
\bu(\bx)\approx \bu(\bx_o)+ \grad_R\!\left\{\bu\!\left(\bx\right)\right\} \left(\bx-\bx_o\right) \eqno{(\theequation{\mathit{a},\mathit{b}})}
\end{equation*}
as can be seen by considering the matrix representation of the two gradient tensors.    This ambiguity was recently examined by  Wood et al.~\cite{wood22-arxiv}, who point out that the question is related to an understanding of the outer product together with the definition of the symbol $\bm\nabla$, as we shall see shortly. 

As with multiplications, there are three different types of first-order derivatives of a vector field, each corresponding to one of the operators ``$\cdot$'', ``$\times$'', and ``$\otimes$'',
\begin{equation}
\bm\nabla\cdot \bu,\quad\quad\quad \bm\nabla\times \bu,\quad\quad\quad \bm\nabla\otimes\bu,\label{threegrads}
\end{equation}
yielding again a scalar, a vector\footnote{More precisely, an axial vector or pseudovector.}, and a tensor, respectively.   Here the $\bm\nabla$ symbol is defined for a Cartesian coordinate system as 
\begin{equation}
\bm\nabla \equiv 
\be^i \frac{\partial}{\partial x^i} \label{deldef}
\end{equation}
where we used the boldface symbol $\bm\nabla$, rather than the more standard $\nabla$, to emphasize its vector-valued nature and to distinguish it from the covariant derivative encountered subsequently.  In the summation convention an upper subscript in the denominator is regarded as a lower subscript, and thus (\ref{deldef}) is equivalent to $\bm\nabla \equiv \sum_{i=1}^3\be_i\frac{\partial}{\partial x^i}$, recalling that $\be^i=\be_i$ in our Cartesian coordinate system.  This definition together with those of the inner, cross, and outer product unambiguously define the three types of derivatives in (\ref{threegrads}) as
\begin{equation}
\bm\nabla\cdot \bu =    \be^i\cdot \frac{\partial \bu}{\partial x^i},\quad\quad\quad
\bm\nabla\times\bu =   \be^i\times \frac{\partial \bu}{\partial x^i},\quad\quad\quad
\bm\nabla\otimes \bu=  \be^i\otimes \frac{\partial \bu}{\partial x^i}. 
\label{threegradsdefs}
\end{equation}
Note that there is no transposition ambiguity in these quantities provided one begins with the definition of $\bm\nabla$ in (\ref{deldef}). We then see that $\bm\nabla\otimes\bu=\grad_L\!\left\{\bu\right\}$, since  the $i$th row and $j$th column of the matrix $[\bm\nabla\otimes\bu]_S$ is found to be $\left[\be^i\right]_S^T\left[ \bm\nabla\otimes\bu\right]_S\left[\be^j\right]_S= \partial u^j/\partial x^i $, matching  (\ref{graddefs}a).  The outer product $\bm\nabla\otimes \bu$ contains within itself the other two derivatives, as one finds 
\begin{equation}\label{twonablaidentities}
\tr\left\{\bm\nabla\otimes \bu\right\}  = \bm\nabla \cdot \bu,\quad\quad\quad
\mskew\!\left\{ \nabla \otimes\bu \right\} =-  \frac{1}{2} \left(\nabla\times \bu\right)^\times
\end{equation}
which follow from $\tr\left\{\ba \otimes \bb\right\}  = \ba \cdot \bb$ in (\ref{dyadtrace})  and $\mskew\!\left\{ \ba\otimes\bb \right\} = -  \frac{1}{2} \left(\ba \times \bb\right)^\times$ in (\ref{skewcorrespond}), respectively.  To see the first identity, note   $\tr\left\{\ba \otimes \bb\right\}  = \ba \cdot \bb$  implies  $\tr\left\{\bm\nabla\otimes \bu\right\}=\frac{\partial}{\partial x^i}\tr\left\{\be^i\otimes \bu\right\}=\frac{\partial}{\partial x^i} \be^i\cdot \bu =\frac{\partial}{\partial x^i}u^i=\bm\nabla\cdot \bu$. The derivation of the second identity beginning with  $\mskew\!\left\{ \ba\otimes\bb \right\} = -  \frac{1}{2} \left(\ba \times \bb\right)^\times$ proceeds similarly.
 
Clearly the first two operations in  (\ref{threegradsdefs}) are the divergence and the curl, and it therefore seems quite natural to refer to the third as the gradient.  By this reasoning, $\bm\nabla\otimes\bu=\grad_L\!\left\{\bu\right\}$ would be called the gradient of $\bu$, with $\grad_R\!\left\{\bu\right\}$  identified as its transpose.  However, in the literature one finds a diversity of opinion as to whether $\bm\nabla\otimes\bu$ should be associated with the gradient of $\bu$.  Irgens  [\citen{irgens}, (4.1.48)] defines the gradient as $\grad_R\!\left\{\bu\right\}$, but then equates $\bm\nabla\otimes\bu$ as the transpose of this leading again to  $\bm\nabla\otimes\bu=\grad_L\!\left\{\bu\right\}$.   Murakami \cite[(12.156)]{murakami} impartially recognizes \emph{both} $\grad_L\!\left\{\bu\right\}=\bm\nabla\otimes \bu$ and $\grad_R\!\left\{\bu\right\}=\bu\otimes \bm\nabla$ to be the gradient of $\bu$, discriminated by whether  $\bm\nabla$ operates to its right or to its left.  Ogden [\citen{ogden}, (1.5.3)]  defines the gradient as being the same as $\bm\nabla\otimes\bu$, yet then defines the latter via the directional derivative in (\ref{graddefsdirectional}b); the result is to set $\bm\nabla\otimes\bu=\grad_R\!\left\{\bu\!\left(\bx\right)\right\}$, a choice that is inconsistent with the definition of $\bm\nabla$ in (\ref{deldef}).   We advocate for identifying $\bm\nabla\otimes \bu$, and not its transpose,  as the tensor-valued gradient of $\bu$, consistent with the definition of the $\bm\nabla$ symbol in (\ref{deldef}) and the natural definitions of divergence, curl, and gradient in (\ref{threegradsdefs}).

The multiplicity of views surrounding the definition of the gradient may partly be ascribed to notational conventions.  Whereas for multiplications one traditionally writes $ab$ for scalars and $\ba\cdot\bb$, $\ba\times\bb$, and $\ba\otimes\bb$ for vectors, for spatial derivatives one writes $\bm\nabla \varphi$ for a scalar field but $\bm\nabla\cdot\bu$, $\bm\nabla\times \bu$ and again $\bm\nabla \bu$ for a vector field.  As pointed out by Wood et al.~\cite{wood22-arxiv}, and emphasized again here, $\bm\nabla \bu$ must be understood as representing  the outer product, which was clearly Gibbs' intent, see for example \cite[p.~66]{gibbs}.  In Gibbs' own notation, what we refer to as the outer product $\ba\otimes \bb$ is represented simply as the juxtaposition $\ba\bb$, and thus $\bm\nabla\bu$ means to Gibbs the outer product (or dyad, in his original terminology) between $\bm\nabla$ and $\bu$.  It seems that in the intervening years between 1884 and the present there have been two developments that have muddled the interpretation of $\bm\nabla \bu$: firstly, the notion of an outer product or dyad construction is frequently omitted from elementary vector analysis, and secondly, when they are used, outer products are typically represented explicitly with the ``$\otimes$'' symbol and not as the  juxtaposition $\ba\bb$. Yet the change in notation for outer products from $\ba\bb$ to $\ba\otimes\bb$ has not led to a comparable change in notation from  $\bm\nabla\bu$ to $\bm\nabla\otimes\bu$---possibly  because the original meaning of $\bm\nabla \bu$ as an outer product was somehow forgotten.

This discussion highlights the fact that $\bm\nabla\otimes \bu$ and not $\bm\nabla \bu$ is a more sensible modern notation for the tensor-valued gradient of a vector field.  As an example, consider the nonlinear  term $\left(\bu\cdot\bm\nabla\right)\bu$ that is ubiquitous in fluid mechanics, representing the advection of momentum by the velocity field $\bu$ itself.  From (\ref{oprodformright}) for the operation of an outer product from the right, we have at once $\left(\bu\cdot\bm\nabla\right)\bu=\bu \left(\bm\nabla\otimes \bu\right)$, and thus we can  express the advection term as the result of the velocity gradient tensor $\bm\nabla \otimes \bu$ operating on the velocity vector $\bu$ to its left.  This is not evident without appreciating the fact that $\bm\nabla\otimes \bu$ is an outer product. 

 
\subsection{Curvilinear coordinates*} \label{curvilinear}

At this point we expand our treatment to include more general curvilinear coordinates, still in ordinary three-dimensional  space, following the presentation in works such as \cite{itskov,dimitrienko,borisenko,ogden,grinfeld}.  With $x^i$ continuing to represent Cartesian coordinates and with $\be_i$ being the  associated orthogonal basis, we will let $q^i$ denote arbitrary curvilinear coordinates with basis vectors $\bg_i$ and  reciprocal basis vectors $\bg^i$ defined such that $\bg_i\cdot \bg^j =\delta_i^j$ where $\delta_i^j$ is the Kronecker delta function.  Furthermore we assume that the coordinate system is right-handed such that the scalar-valued triple product of the basis vectors is positive, $\left(\bg_1\times\bg_2\right)\cdot \bg_3>0$, and we introduce the volume element $\sqrt{g}\equiv \left(\bg_1\times\bg_2\right)\cdot \bg_3$. Note that in curvilinear coordinate systems the basis vectors and the reciprocal basis vectors are in general both spatially varying.

It is worth taking a moment to clarify the nature of the reciprocal basis $\bg^i$. Like the basis vectors $\bg_i$, the reciprocal basis vectors $\bg^i$ are proper vectors in the sense of physics, that is, quantities that are characterized by a magnitude and direction which we can visualize as arrows in space, and both  are bases for the usual three-dimensional Euclidian vector space.   In more general classes of spaces for which no inner product is defined, it may be more useful to consider instead a set of vectors $\bg_i$ and their dual vectors $\bg^i_*$.  These dual vectors represent linear mappings of vectors to real numbers.  While no longer being vectors in the sense of physics, these mappings remain vectors in the mathematical sense of being elements of a vector space, in this case the dual vector space to the space inhabited by the vectors.  This is the orientation adopted in differential geometry \cite{hicks,burke,marsden,bishop,spivak,needham,carroll}, an elegant and powerful framework for the study of abstract surfaces and volumes that greatly generalizes our normal intuitive notions of these quantities, but which  requires considerable study to master.  An enlightening discussion of the relationship between the reciprocal basis and the dual basis may be found in \S1.4 of Odgen \cite{ogden}, who ultimately concludes that for  three-dimensional space in which an inner product is readily available, the use of the reciprocal basis is perfectly sufficient. 
  
In a general curvilinear coordinate system, one can write a vector equivalently in terms of the basis or reciprocal basis as $\ba=a^i \bg_i=a_i\bg^i$.  Here the coefficients $a^i\equiv \ba \cdot \bg^i$ and $a_i\equiv \ba \cdot \bg_i$ are called the \emph{contravariant} and \emph{covariant} components of $\ba$, respectively, owing to their transformation laws under a change in the coordinate system.  In this notation the inner and cross products become  \cite[(1.29) \& (1.45)]{itskov}
\begin{equation}
\ba\cdot\bb = a^ib_i = a_ib^i,\quad\quad\quad\ba\times\bb = \varepsilon_{ijk} a^i b^j \bg^k = \varepsilon^{ijk} a_i b_j \bg_k  \label{curvdotandcross}
\end{equation}
where  $\varepsilon_{ijk}$ and $\varepsilon^{ijk}$ are the components of the Levi-Civita tensor, defined in terms of the  Levi-Civita symbol  $\epsilon^{ijk}=\epsilon_{ijk}$---which is equal to $+1$, $-1$, or $0$ for  $(i,j,k)$ being respectively an even, odd, or other permutation of $(1,2,3)$---as $\varepsilon_{ijk}=\epsilon_{ijk} \sqrt{g}$ and $\varepsilon^{ijk}=\epsilon^{ijk}/\sqrt{g}$ respectively, see \S 2.8 of \cite{carroll} or \S 9 of \cite{grinfeld}.  
 Similarly, the outer product is given by
 \begin{equation}
 \ba\otimes \bb  = a^ib^j \bg_i\otimes \bg_j  = a_ib^j \bg^i\otimes \bg_j  = a^ib_j \bg_i\otimes \bg^j =a_ib_j \bg^i\otimes \bg^j 
\end{equation}
and we note that any second-order tensor can be written equivalently as four sums of nine outer products as
\begin{equation}
\bM = M^{ij} \bg_i\otimes \bg_j  = M_i^{\cdot j} \bg^i\otimes \bg_j    = M^i_{\cdot j} \bg_i\otimes \bg^j  = M_{ij} \bg^i\otimes \bg^j
\end{equation}
where the coefficients are defined as $M^i_{\cdot j}\equiv  \bg^i \bM \bg_j$, etc.   For the special case of an orthonormal basis, the basis vectors and reciprocal basis vectors are identical.  Thus in a Cartesian coordinate system, in which the basis vectors are denoted $\be_i$, we therefore have $\be^i=\be_i$, giving meaning to a statement that was previously introduced merely as notation. Consequently, $a_i=a^i$,  $b_i=b^i$,  and  $M^{ij} = M_i^{\cdot j}=M^i_{\cdot j}= M_{ij}$, and  the above expressions reduce to the those given previously for a Cartesian coordinate system.   

One distinguishing feature of the index notation in comparison with symbolic notation is that there is no explicit notion of a transpose operation in the former, nor of left operation, as discussed in \S 7.2 of Grinfeld \cite{grinfeld}.  Instead, in index notation determining the suitable representations for the tensor and vector elements is accomplished by index raising and lowering operations, using operations with the identity or metric tensor $\bI =g^{ij} \bg_i\otimes \bg_j  = \bg^i\otimes \bg_i  =  \bg_i\otimes \bg^i  =g_{ij} \bg^i\otimes \bg^j $ if required; here $g^{ij}\equiv \bg^i\bI\bg^j=\bg^i\cdot \bg^j$ and $g_{ij}\equiv \bg_i\bI\bg_j=\bg_i\cdot \bg_j$.  As the identity tensor is the tensor whose action does nothing, we can use it to change representations without modifying the underlying proper vector. 

In tensor analysis it is commonplace to refer to $a^i$ and $a_i$ as contravariant and covariant \emph{vectors}.  It is crucial to understand that this is a shorthand, as discussed e.g. in Carroll \cite{carroll} on p. 17 and p. 22.  Strictly speaking,  $a^i$ and $a_i$ are the contravariant and covariant components representing the same proper vector $\ba$, which can equivalently be represented either as $\ba=a^i\bg_i$ or $\ba=a_i\bg^i$, utilizing the basis or reciprocal basis respectively.  The proper vector does not change when we choose to represent it in a different basis.  Similarly, referring to $M^{ij}$, $M_{ij}$, and $M_i^{\cdot j}$ and $M^i_{\cdot j}$ as contravariant, covariant, and mixed tensors, respectively, is also a shorthand.  These conventions reflect the fact that the basis vectors can be unambiguously inferred from the positions of the indices on the components; thus, since we know that $M^i_{\cdot j}$ must be associated with $\bg_i\otimes\bg^j$, we can refer to  $M^i_{\cdot j}$ as the tensor and understand that the basis vectors are implied.  This allows for a remarkable compactness of notation, as we can then say, for example, that a new vector $c^i$ is defined as $c^i\equiv M^i_{\cdot j} a^j$, understanding this to be a shorthand for $\bc=c^i\bg_i \equiv M^i_{\cdot j} \bg_i\otimes\bg^j a^k \bg_k=M^i_{\cdot j} a^j \bg_i$.  More generally, if the indices to be summed over are arranged in upper and lower pairs in accord with the summation convention, the basis vectors follow their components and can generally be ignored---or rather, implied---in correctly formed operations.  At the same time, this compactness can blur the distinction between proper vectors or tensors and their components; if the basis vectors are invisible, one must take care to ensure that they are not entirely forgotten.   Herein, basis vectors will always be written explicitly when the vector or tensor, and not its components, are intended.

In curvilinear coordinates a complication arises when taking spatial derivatives due to the fact that the basis vectors themselves vary spatially.  To accommodate this, with $\partial_i\equiv \partial/\partial q^i$ being a shorthand for the partial derivative with respect to the $i$th coordinate, we write  the derivatives of a vector $\ba$ as
\begin{align}
\partial_i\ba &= \frac{\partial}{\partial q^i}\left(a^j \bg_j\right)=\left(\partial_i a^j\right) \bg_j +  a^j\partial_i \bg_j  = \left( \partial_i a^j  +a^k \Gamma^j_{ik}\right)\bg_j,& \partial_i\bg_j &= \Gamma_{ij}^k \bg_k \label{a1expand}\\
\partial_i \ba  &= \frac{\partial}{\partial q^i}\left(a_j \bg^j\right)=\left(\partial_i a_j\right) \bg^j +  a_j\partial_i\bg^j  = \left( \partial_i a_j -a_k \Gamma^k_{ij}\right)\bg^i,  & \partial_i \bg^j  &= -\Gamma_{ik}^j \bg^k \label{a2expand}
\end{align}
when expressed in terms of the basis or the reciprocal basis, see  (2.81--2.82) and (2.91--2.92) of \cite{itskov}; note that the $j$ and $k$ indices have been swapped at the final equality.  The key step here is expanding the derivatives of the basis vectors in terms of the basis vectors themselves, leading to coefficients  $\Gamma_{ij}^k$, called the Christoffel symbol of the second kind, that are symmetric in their lower indices and that are given by  [\citen{itskov}, (2.80--2.81)]
\begin{equation}
\Gamma_{ij}^k=\Gamma_{ji}^k\equiv  \left( \partial_j \bg_i\right) \cdot \bg^k =\left( \partial_i \bg_{j}\right) \cdot \bg^k=-\left( \partial_j \bg^k\right) \cdot \bg_i=-\left( \partial_i \bg^k\right) \cdot \bg_j\label{gammadef}
\end{equation}
with the latter two equalities are obtained by differentiating the orthogonality relation $\bg^i\cdot\bg_j=\delta^i_j$. The two equalities on the right-hand side of (\ref{a1expand}) and (\ref{a2expand}) follow from this definition.

The \emph{covariant derivative} with respect to the $i$th coordinate, denoted $\nabla_i$, is then defined as the coefficient of the basis vector in the partial derivative expansions (\ref{a1expand}) and (\ref{a2expand}) 
\begin{align}
\partial_i \ba & = \left(\nabla_i a^j \right) \bg_j, \quad\quad\quad\quad \nabla_i a^j  \equiv  \partial_i a^j +a^k \Gamma^j_{ik}\label{dgamma1}\\
\partial_i \ba &= \left( \nabla_i a_j \right) \bg^j,\quad \quad\quad\quad\nabla_i a_j  \equiv    \partial_i a_j -a_k \Gamma^k_{ij}.\label{dgamma2}
\end{align}
see (2.72) and (2.93) of \cite{itskov}.   Note that the notation  $\nabla_i$ for the covariant derivative, which is conventional \cite{carroll,grinfeld}, does not mean the $i$th component of $\bm\nabla$, yet in a sense behaves as if it were;  $\nabla_i$  will be found to appear in locations in which one would expect the $i$th component of $\bm\nabla$ if the latter were an ordinary vector. In Cartesian coordinates, the basis vectors are constant, so the Christoffel symbols vanish and $\nabla_i$ reduces to simply the $i$th partial derivative $\partial_i$.  An important property of the covariant derivative is that it vanishes when applied to the basis and reciprocal basis, the so-called \emph{metrinilic property}, see the discussion in \S 8.6.7 of Grinfeld \cite{grinfeld}.  That is, we have
\begin{equation}
\nabla_i \bg^j   = \partial_i \bg^j +\bg^k \Gamma^j_{ik}=\bm{0},\quad\quad\quad
\nabla_i \bg_j =    \partial_i \bg_j -\bg_k \Gamma^k_{ij}=\bm{0}\label{nilic}
\end{equation}
as is found by taking the dot product of the left equation with $\bg_\ell$, and of the right equation with $\bg^\ell$, followed by the use of the orthogonality condition together with the expression for $\Gamma_{ij}^k$ in (\ref{gammadef}). 


Although it does not frequently appear in the literature, a consistent definition of $\bm\nabla$ in curvilinear coordinates is given by [\citen{dimitrienko}, (2.23)\footnote{Note however a typographic error in  (2.23) of \cite{dimitrienko}, which incorrectly includes an outer product symbol $\otimes$ in the definition of $\bm\nabla$ itself.}; \citen{irgens}, (12.5.20)]
\begin{equation}
\bm\nabla \equiv  \bg^i \frac{\partial}{\partial q^i} \label{deldefoblique}
\end{equation}
with the three types of derivatives in (\ref{threegrads})  then being given by, see e.g \S5.1.5 of \cite{borisenko} or \S2.1.5 of \cite{dimitrienko},
\begin{equation}
\bm\nabla\cdot \bu =   \bg^i\cdot \partial_i \bu,\quad\quad\quad
\bm\nabla\times\bu = \bg^i\times\partial_i \bu,\quad\quad\quad
\bm\nabla\otimes \bu= \bg^i\otimes \partial_i \bu.  
\label{threegradsdefscurv}
\end{equation}
In terms of the covariant derivatives, these become 
\begin{equation}
\bm\nabla\cdot \bu =    
 \nabla_i u^i\,,\quad\quad\quad
\bm\nabla\times\bu =  \left(\nabla_i u_j  \right) \bg^i\times \bg^j =
\varepsilon^{ijk}\left( \partial_i u_j\right)\bg_k ,\quad\quad\quad
\bm\nabla\otimes \bu=
\left(\nabla_i u^j \right)\bg^i\otimes  \bg_j
\label{threegradsdefscurv}
\end{equation}
after substituting (\ref{dgamma1}) into the first and third expressions and  (\ref{dgamma2}) into the second, making use of the fact that we have $\bg^i\times \bg^j = \varepsilon^{ijk} \bg_k$ in three dimensions \cite[(1.44)]{itskov}. Note that in the expression for the curl, the covariant derivative $\nabla_i u_j$ may be replaced  with the partial derivative\footnote{To see this, consider the case $k=3$ for concreteness.  Then $\varepsilon^{ij3}\left( \nabla_i u_j\right)\bg_3$ involves a summation over nine $(i,j,3)$ terms, only two of which---$(1,2,3)$ and $(2,1,3)$---are nonzero.  Then from (\ref{dgamma2}) we have \[\varepsilon^{ij3}\!\left( \nabla_i u_j\right)\bg_3=\frac{1}{\sqrt{g}}\left(\nabla_1u_2-\nabla_2 u_1\right)\bg_k=\frac{1}{\sqrt{g}}\left[\left(\partial_1 u_2 -u_k\Gamma_{12}^k\right)-\left(\partial_2 u_1 -u_k\Gamma_{21}^k\right)\right]\bg_k=\frac{1}{\sqrt{g}}\left(\partial_i u_j-\partial_i u_j\right)\bg_3=\varepsilon^{ij3}\!\left( \partial_i u_j\right)\bg_k\] by the symmetry of the Christoffel symbol, and similarly for $k=1$ and $k=2$. This verifies that $\varepsilon^{ijk}\left( \nabla_i u_j\right)\bg_k=\varepsilon^{ijk}\left( \partial_i u_j\right)\bg_k$.}  $ \partial_i u_j$ on account of the symmetry $\Gamma_{ij}^k=\Gamma_{ji}^k$ together with the properties of $\varepsilon^{ijk}$.  The result for the outer product matches Dimitrienko \cite{dimitrienko}, see (2.24) therein, and Irgens \cite{irgens}, see their (12.5.35) together with (4.4.18), but differs by a transpose from (1.5.14) of  Ogden \cite{ogden} for the reasons discussed in the previous section.    These three equations all reduce to their familiar forms in the Cartesian case, for which  all covariant derivatives can be replaced with partials.  

As an aside, we point out one reason we have chosen the notation  $\nabla_i u^j$ for the covariant derivative.  As mentioned previously, in index notation when one writes $M_{ij}$ this is generally understood to mean the entire tensor including the basis vectors, $M_{ij}\bg^i\otimes\bg^j$, and not simply a single component.  In particular, $M_i^{\cdot j}$ means $M_i^{\cdot j}\bg^i\otimes\bg_j$ and not $M_i^{\cdot j}\bg_j\otimes\bg^i$ because, although the latter does have the correct mix of contravariant and covariant indices, by convention we write the basis vectors with indices in the order in which they appear in the tensor coefficient; the dot in $M_i^{\cdot j}$ plays the important role of a placeholder that specifies $i$ as the first index and $j$ as the second. The notations $u^{j}|_i$ and $u^{j}_{;i}$ that one also sometimes sees for the covariant derivative have their indices in the wrong order for correctly inferring the ordering of the basis vectors as $\bg^i\otimes  \bg_j $, which itself is fixed by our choice of the outer product $\bm\nabla\otimes\bu$ as being set by the definition of $\bm \nabla$.  That the basis vector corresponding to the derivative appears in the first position in the outer product comprising the tensor is one factor in favor of a derivative symbol that appears before the quantity being differentiated.  

As pointed out by e.g. Borisenko and Tarapov  \cite[(4.43)]{borisenko} and Murakami  \cite[(12.170)]{murakami}, the divergence, curl, and gradient tensor theorems can be expressed in the unified form 
\begin{equation}
\bm\nabla \ast (\cdot) = \lim_{V\longrightarrow 0}\frac{1}{V}\iint_\calA  \bn \ast  (\cdot)  \,  \rd A\label{borisenkovardef}
\end{equation}
where $\ast$ is one of $\cdot$, $\times$, or $\otimes$, and the operand $(\cdot)$ could be a scalar, vector, or tensor field.  The four cases $\bm\nabla \varphi$, $\bm\nabla \cdot \bu$, $\bm\nabla \times \bu$, and $\bm\nabla \otimes \bu$ follow from taking the limits of the presented subsequently  three-dimensional gradient (\ref{borisenkotheorem}), divergence (\ref{divergencetheorem}), curl (\ref{ith3dtensortheorem}), and gradient tensor (\ref{3dtensortheoremnew}) theorems, respectively.   Borisenko and Tarapov \cite{borisenko} propose this equation as a coordinate-independent definition of $\bm\nabla$, from which the Cartesian representation   $\bm\nabla=  \be^i \frac{\partial}{\partial x^i} $, given earlier in (\ref{deldef}) as a definition, would follow as a consequence. However we prefer instead to see  $\bm\nabla \equiv  \bg^i \frac{\partial}{\partial q^i}$ in (\ref{deldefoblique}) as the coordinate-independent definition of  $\bm\nabla$, with (\ref{borisenkovardef}) then being  a consequence.  

\section{Results}\label{derivation}

We now turn to the gradient tensor theorem itself, examining it in several different ways. After presenting a proof based on the divergence theorem, we show that the gradient tensor theorem contains both the divergence theorem as well as a three-dimensional generalization of the Kelvin--Stokes theorem.  Then, we decompose the two-dimensional gradient tensor with a set of basis tensors, and show that the nonstandard portions of this identity involve what would be interpreted as straining terms for the case  that our vector field corresponds to a  velocity.  Finally, we ask whether the two-dimensional gradient tensor theorem also holds for curved surfaces, and find that it does not.

 \subsection{Derivation}

As mentioned in the introduction, in Gibbs' 1884 monograph \cite{gibbs} in which he lays out the tools of modern vector analysis, the following identity appears [eqn. (2), p.~65, \S161]
\refstepcounter{equation}
\begin{equation*}
\iiint_\calV  \bm\nabla \otimes \bu\, \rd V=\iint_\calA  \bn \otimes \bu\, \rd \quad\quad\quad\bigg|\quad\quad\quad\iiint_\calV \nabla_iu^j \bg^i\otimes \bg_j \, \rd V=\iint_\calA  n_i u^j  \bg^i\otimes \bg_j \, \rd A
\eqno{(\theequation{\mathit{a},\mathit{b}})}\label{3dtensortheoremnew}
\end{equation*}
which we refer to as the \emph{gradient tensor theorem}.  This has been written in both symbolic and index notation, with symbolic notation  on the left and index notation on the right, separated by a vertical line.  This convention will be followed in most equations in this and the following subsections, allowing readers to choose whichever version they prefer, with the index notation versions employing a curvilinear coordinate system for generality.  Here the index notation version of the gradient tensor theorem follows from its symbolic version using  the  definitions of the outer product and the $\bm\nabla$ operator in curvilinear coordinates given in \S \ref{prelim}\ref{curvilinear}. 

As in Gibbs' work, the gradient tensor theorem can be derived for a cuboidal volume aligned with the coordinate axes by integrating each term using the fundamental theorem of calculus.  More generally, it can be proved very simply beginning with the usual three-dimensional divergence theorem,
\refstepcounter{equation}
\begin{equation*}
\iiint_\calV  \bm\nabla \cdot \bu\, \rd V=\iint_\calA  \bn \cdot \bu\, \rd A\quad\quad\quad\bigg|\quad\quad\quad\iiint_\calV \nabla_iu^i\, \rd V=\iint_\calA  n_i u^i  \, \rd A.
 \eqno{(\theequation{\mathit{a},\mathit{b}})} \label{divergencetheorem}
\end{equation*}
Following \S4.3.4 of Borisenko and Taparov \cite{borisenko}, consider a vector field of the particular form 
\refstepcounter{equation}
\begin{equation*}
\bu(\bx) = \bc \varphi(\bx) \quad\quad\quad\bigg|\quad\quad\quad u^i(\bx) =  c^{i}(\bx) \varphi(\bx) \eqno{(\theequation{\mathit{a},\mathit{b}})}
\end{equation*}
where $\bc=c^i\bg_i$ is an arbitrarily chosen, spatially uniform vector\footnote{While $\bc$ itself is spatially uniform, recall that its components in curvilinear coordinates may nevertheless be a function of space.} and $\varphi(\bx)$ is a scalar field.  From the definition of the divergence in (\ref{threegradsdefs}) we have the chain rule for divergence, $\bm\nabla\cdot(\varphi\bu) = \varphi\, \bm\nabla \cdot \bu + \bu \cdot \bm\nabla \varphi$, which for a spatially uniform vector $\bc$ simplifies to $\bm\nabla\cdot(\varphi\bc) =  \bc \cdot \bm\nabla \varphi$. 
The divergence theorem  yields
\refstepcounter{equation}
\begin{equation*}
\bc \cdot \iiint_\calV  \bm\nabla \varphi \, \rd V= \bc \cdot \iint_\calA \bn\varphi\, \rd A\quad\quad\bigg|\quad\quad  c^{j}\bg_j \cdot  \iiint_\calV \left( \partial_i \varphi\right)\bg^i \rd V=c^{j}\bg_j \cdot \iint_\calA   n_i \varphi  \bg^i\rd A
 \eqno{(\theequation{\mathit{a},\mathit{b}})} \label{borisenkotheoremintermediate}
\end{equation*}
which, since $\bc$ can be any vector, implies an identity given as  (4.30) in \cite{borisenko} and  as (1) in  \S 161 of Gibbs \cite{gibbs}
\refstepcounter{equation}
\begin{equation*}
\iiint_\calV  \bm\nabla \varphi\, \rd V=\iint_\calA \bn\varphi\, \rd A\quad\quad\quad\bigg|\quad\quad\quad   \iiint_\calV \left( \partial_i \varphi\right)\bg^i \rd V= \iint_\calA    n_i \varphi   \bg^i  \rd A.
 \eqno{(\theequation{\mathit{a},\mathit{b}})} \label{borisenkotheorem}
\end{equation*}
This result, which we refer to as the \emph{gradient theorem for volume integrals}, states that the integral of the gradient of a scalar field $\bm\nabla \varphi$ over a volume is equal to the integral of the normal vector $\bn$, weighted by the value of the scalar field $\varphi$, over the bounding surface.   Now, letting both sides of (\ref{3dtensortheoremnew}) operate to their right on $j$th Cartesian reciprocal vector $\be^j$,  and for clarity denoting $j$th component of $\bu$ specifically in a Cartesian coordinate system as $ \breve u^j\equiv \bu\cdot \be^j$, we note that $\left(\bm\nabla\otimes \bu\right) \be^j= \bm\nabla \breve u^j$ and $(\bn \otimes \bu) \be^j= \bn \breve u^j$ leading to
\refstepcounter{equation}
\begin{equation*}
\iiint_\calV  \bm\nabla \breve u^j\, \rd V=\iint_\calA   \bn\breve u^j\, \rd A\quad\quad\quad\bigg|\quad\quad\quad  \iiint_\calV \left(\partial_i \breve u^j\right)\bg^i  \,\rd V= \iint_\calA n_i \breve u^j \bg^i    \rd A 
 \eqno{(\theequation{\mathit{a},\mathit{b}})} \label{borisenkotheoremforu}
\end{equation*}
which is simply (\ref{borisenkotheorem}) with $\varphi$ replaced with $\breve u^j$.  Note carefully the unusual construction in the right column: the gradient operator and normal vector are still represented with a general curvilinear basis, but the component of $\bu$ involved is its $j$th Cartesian component.  The gradient tensor theorem is thus seen to be an aggregation of three versions of the gradient theorem for volume integrals, one for each Cartesian component of $\bu$. 

Equivalently, since in Euclidean space the Cartesian coordinate system is always available,  we can choose the Cartesian basis $\be_i$ to represent our vectors and find
\begin{equation}
\iiint_\calV  \bm\nabla \otimes \bu\, \rd V =\iiint_\calV  \be^i \otimes  \ \frac{\partial}{\partial x^i} \left(u^j\be_j\right) \rd V=\be^i \otimes \be_j\iiint_\calV  \frac{\partial}{\partial x^i} u^j \rd V \label{alternate1}
\end{equation}
for the  left-hand-side of the gradient tensor theorem. Here the fact that the basis vectors are spatially constant implies then $\be_j$ can be pulled outside of the derivative and  that the outer product $\be^i \otimes \be_j$ can then be pulled outside of the integral.  The right-hand side is similarly found to be
\begin{equation}
\iint_\calA  \bn \otimes \bu\, \rd A=\iint_\calA  \left(n_i \be^i\right) \otimes\left( u^j \be_j\right) \, \rd A= \be^i \otimes \be_j \iint_\calA  n_i u^j  \rd A \label{alternate2}
\end{equation}
but we see that the final integrals in (\ref{alternate1}) and (\ref{alternate2}) are the $i$th components of those in (\ref{borisenkotheoremforu}), showing again that the gradient tensor theorem is true by the gradient theorem for volume integrals. 

The use of the Cartesian basis is crucial here.  Since the basis vectors $\bg_i$ and $\bg^i$ appearing in (\ref{3dtensortheoremnew}b) are in general spatially varying, we cannot pull the  outer products $\bg^i\otimes \bg_j$ outside the integrals, and similarly we also cannot isolate a component by taking the inner product of both sides with $\bg_k$ from the left and $\bg^\ell$ from the right. Indeed, we need not even have chosen the same curvilinear coordinates and basis vectors on the two sides.  Thus if we attempt to omit the basis vectors from (\ref{3dtensortheoremnew}b), we obtain 
\begin{equation}
\iiint_\calV \nabla_iu^j  \rd V\ne \iint_\calA  n_i u^j   \rd A  \label{nottrue}
 \end{equation}
which involves integrals over tensor components and not proper tensors themselves; importantly, the equality does not hold in general curvilinear coordinates.  On the other hand, if we  choose the Cartesian basis $\be_i$ for our representations, then in this case we can pull the  outer products $\be^i\otimes \be_j$ outside the integrals as in (\ref{alternate1}) and (\ref{alternate2}), and in that case the equality in the above expression is obtained;  the same applies for (\ref{borisenkotheoremforu}b), which will become the same equality in a Cartesian coordinate system.  This situation, where the Cartesian and curvilinear basis vectors must be treated differently with respect to integration, illustrates the potential confusion arising from a notation in which the basis vectors are implicit rather than explicitly specified.

\subsection{A hierarchy of integral theorems}

The gradient tensor theorem contains within itself a number of important special cases.    Taking the trace of the both sides of the gradient tensor theorem, we observe that since the trace is a linear operator it can be brought inside the integrals, leading to  the divergence theorem (\ref{divergencetheorem}). To show this, we simply apply $\tr\left\{\bm\nabla\otimes \bu\right\}  = \bm\nabla \cdot \bu$  together with  $\tr\left\{\ba \otimes \bb\right\}  = \ba \cdot \bb$, given earlier in (\ref{twonablaidentities}) and (\ref{dyadtrace}), respectively, to either the symbolic or index notation form of the gradient tensor theorem in (\ref{3dtensortheoremnew}).  

A different reduced identity is contained within the skew-symmetric part of the gradient tensor theorem, namely the  vector-valued identity 
\refstepcounter{equation}
\begin{equation*}
\iiint_\calV   \bm\nabla \times \bu \,\rd V=\iint_\calA   \bn \times \bu\,  \rd A
\quad\quad\bigg|\quad\quad
\iiint_\calV  \varepsilon^{ijk} \left(\partial_i u_j \right)\bg_k  \,\rd V=\iint_\calA   \varepsilon^{ijk} \,n_i u_j \bg_k\,  \rd A
 \eqno{(\theequation{\mathit{a},\mathit{b}})}\label{ith3dtensortheorem}
\end{equation*}
which can be seen as a three-dimensional analogue of the Kelvin--Stokes theorem.  Whereas the Kelvin--Stokes  theorem links an area integral of one component of the curl to a line integral, here we are connecting a volume integral of all three components of the curl to a surface integral.  For a cuboidal volume aligned with the Cartesian coordinate axes, we can visualize this result as being built up from stacks of planes within which the Kelvin--Stokes theorem is applied, in turn, along each of the three orthogonal directions. 

The three-dimensional curl theorem is found from the gradient tensor theorem through a series of linear operations: (i) take the skew-symmetric part of both sides, which may be brought inside the integrals; (ii) use $\mskew\!\left\{ \nabla \otimes\bu \right\} =-  \frac{1}{2} \left(\nabla\times \bu\right)^\times$ from  (\ref{twonablaidentities}) together with $\mskew\!\left\{ \ba\otimes\bb \right\} = -  \frac{1}{2} \left(\ba \times \bb\right)^\times$  from  (\ref{skewcorrespond}) to convert the skew-symmetric parts of the tensors on both sides to skew tensors associated with a vector, and finally (iii) move the skew tensor operator  ``$^\times$'' outside of the integrals and drop it from  both sides to obtain  (\ref{ith3dtensortheorem}). These three  steps can be applied to either to the symbolic  or index notation form of the gradient tensor theorem in (\ref{3dtensortheoremnew}), leading  to  (\ref{ith3dtensortheorem}a) and  (\ref{ith3dtensortheorem}b) respectively.  In the latter case we begin with 
\begin{equation}
\iiint_\calV \nabla_iu_j \bg^i\otimes \bg^j \, \rd V=\iint_\calA  n_i u_j  \bg^i\otimes \bg^j \, \rd A,\label{3dtensortheoremnew2}
\end{equation}
which employs the covariant representation for $\bu$, and then make use of the fact that $\bg^i\times \bg^j=\varepsilon^{ijk} \bg_k$ at  step (ii).  That the two versions of  (\ref{ith3dtensortheorem}) match is  verified by observing that $\bm\nabla \times \bu =  \varepsilon^{ijk} \left(\partial_i u_j \right)\bg_k$  from (\ref{threegradsdefscurv}) together with $\bn\times\bu= \varepsilon^{ijk}n_iu_j\bg_k$ from (\ref{curvdotandcross}).
  
The case of a two-dimensional vector field within a flat surface is particularly important, as it arises frequently in fluid dynamics and other applications.    For a vector field $\bu$ lying within a horizontal surface having vertical unit normal vector $\bk$, we have a two-dimensional version of the gradient tensor theorem
\refstepcounter{equation}
\begin{equation*}
\iint_\calA  \bm\nabla \otimes \bu\, \rd A=\oint_\calL  \rd \bn \otimes \bu 
\quad\quad\quad\bigg|\quad\quad\quad
\iint_\calA \nabla_i u^j \bg^i\otimes \bg_j \, \rd A=\oint_\calL    \bg^i\otimes \bg_j  u^j    \rd n_{i} \eqno{(\theequation{\mathit{a},\mathit{b}})} \label{flowtheoremnew}
\end{equation*}
as well as the familiar two-dimensional divergence  and Kelvin--Stokes theorems
\refstepcounter{equation}
\begin{equation*}
\begin{array}{lcl}\displaystyle  \iint_\calA  \bm\nabla \cdot  \bu\, \rd A=\oint_\calL \bu \cdot \rd\bn &\quad\quad\quad\bigg|\quad\quad\quad & \displaystyle  \iint_\calA \nabla_iu^i\, \rd A=\oint_\calL u^i \rd n_i  \vspace{.1in}\\
\displaystyle   \iint_\calA  \bk\cdot\bm\nabla \times \bu\, \rd A=\oint_\calL \bu \cdot  \rd \bx &\quad\quad\quad\bigg|\quad\quad\quad &   \displaystyle   \iint_\calV  \varepsilon^{ij3} \partial_i u_j \,\rd V=\oint_\calA  \varepsilon^{ij3} n_i u_j \,  \rd A 
\end{array}
  \eqno{(\theequation{\mathit{a},\mathit{b}})}\label{divandstokes}
\end{equation*}
where for the index notation  versions we have chosen the curvilinear coordinate system such that $\bg_3=\bg^3=\bk$.  In the above integrals,  $\rd \bx$ is a differential vector tangent to the boundary $\calL$, and the differential normal vector $\rd\bn$ is defined as $\rd\bn \equiv \bn \,\rd\ell$ with $\bn$ being the exterior normal to the boundary and $\rd\ell$ a differential of arc length along it.  With the boundary integrals being traversed in the right-hand sense, $\rd\bn$ and $\rd\bx$ are related by $\rd\bn=-\bk\times \rd\bx$.  As is the case in three dimensions, the trace  of the 2D gradient tensor theorem is the 2D divergence theorem, while its skew-symmetric part gives the 2D Kelvin--Stokes theorem.   To see the latter, we take the negative of the skew-symmetric part of  (\ref{flowtheoremnew})  to  obtain $\frac{1}{2}\left(\bm\nabla \times \bu\right)^\times$ on the left and $\frac{1}{2}\left(\rd\bn\times\bu\right)^\times$ on the right.  Dropping the ``$^\times$'' operator leads to an equality between vectors, and taking the vertical component gives $\bk \cdot \bm\nabla\times\bu$ on the left and $\bk \cdot \rd \bn \times \bu = \bu \cdot \bk \times \rd \bn=\bu \cdot \rd\bx$  on the right. All three of  these 2D theorems can be derived from their 3D versions by limiting the vector field of interest to a plane. 

\subsection{The 2D gradient tensor theorem for a flat surface}\label{twodgradtensor}

\begin{figure}
\begin{center}
\includegraphics[width=\textwidth,angle=0]{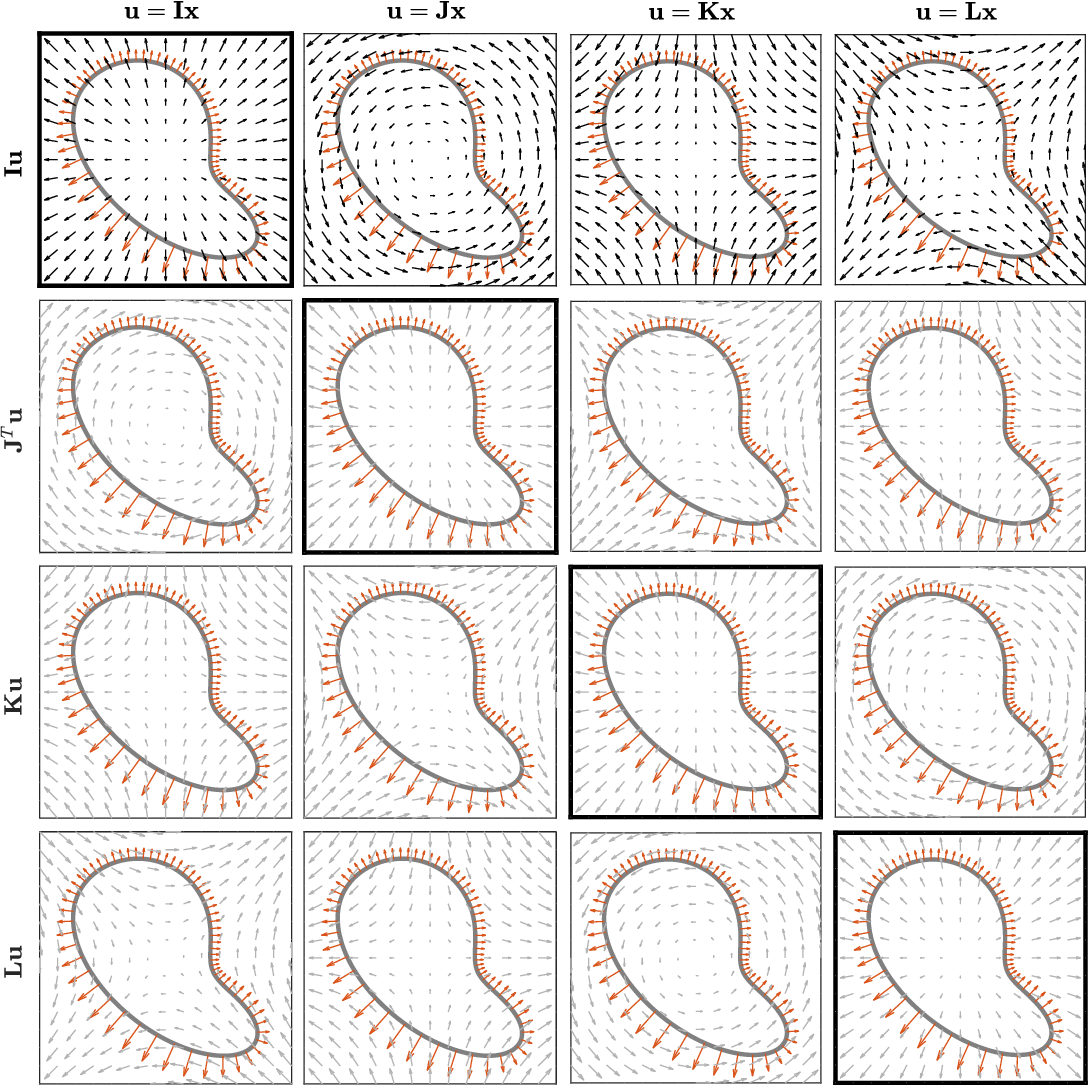}
\end{center}
	\caption{An illustration of the gradient tensor theorem.  The upper row shows vector fields $\bu$ consisting of $\bu=\bI\bx=\bx$,  $\bu=\bJ\bx$,  $\bu=\bK\bx$,  and $\bu=\bL\bx$, respectively, where $\bx$ is the position vector with respect to the origin.  The heavy grey curve in all plots is the bounding curve $\calL$, encompassing a region $\calA$ over which we wish to integrate and shown with its exterior normal vector in orange.  In subsequent rows, the original vector field is multiplied by $\bJ^T$, $\bK$, and $\bL$ respectively; the use of the grey colour for the vectors indicates that this is a modified version of the original field.  These multiplications turn each of the four original fields into each of the other types.  Along the diagonal, marked by the bold bounding boxes, the original vector field is transformed into a purely divergent field.  This illustrates how the gradient tensor theorem can be thought of as the divergence theorem applied to versions of the original vector field modified by the transpose of each of the $\bI\bJ\bK\bL$ tensors. }\label{vectors}
\end{figure}

In this section we examine the gradient tensor theorem in two dimensions.  To do so,  we first define a convenient basis for decomposing 2D tensors. Let our Cartesian coordinate system $S$ now be limited to the horizontal plane, and with a change in notation we will let $x$ and $y$ be the horizontal coordinates, with associated basis vectors $\bi$ and $\bj$ respectively, while $\bk$ will denote the normal vector to the plane.  We write as $\bx=x\bi+y\bj$ for the position vector with respect to the origin and  $\bu=u\bi+v\bj$ for a general horizontal vector~$\bu$. We then define a  set of tensors, termed the $\bI\bJ\bK\bL$ tensors, as 
\begin{equation}
\left. \begin{array}{lll} \displaystyle
\bI  \equiv  \bi \otimes \bi +\bj \otimes \bj&\quad\quad\quad\quad& \displaystyle
\bJ\equiv  \bj \otimes \bi -\bi \otimes \bj\vspace{.1in}\\
 \displaystyle
\!\!\!\bK  \equiv  \bi \otimes \bi -\bj \otimes \bj&&
 \displaystyle
\bL  \equiv  \bj \otimes \bi +\bi \otimes \bj\end{array}\right.
  \label{ijkldef} 
\end{equation}
which are, respectively,  the two-dimensional identity tensor $\bI$, the ninety-degree counterclockwise rotation tensor $\bJ$,  the  reflection tensor $\bK$ about the direction of $\bi$, and the reflection tensor $\bL$ about direction of $\bi+\bj$.  These identifications can be understood by considering how the tensors operate on $\bi$ and $\bj$ using the outer product definition (\ref{oprodform}).  For example, since $\bI\bi=\bi$ and $\bI\bj=\bj$, the action of $\bI\bx$ is to map $\bx$ into itself. Similarly we see that $\bJ\bi=\bj$ and $\bJ\bj=-\bi$, and thus $\bJ\bx$ accomplishes a ninety degree counterclockwise rotation of $\bx$.  As an illustration, the top row of figure~\ref{vectors} shows the vector field arising from each tensor operating on the position vector $\bx$.  The grey contour, and subsequent rows, will be referred to later. 
 
It may be more intuitive to view the $\bI\bJ\bK\bL$ tensors in their matrix  representations.  In our two-dimensional  Cartesian coordinate system $S$ with basis vectors $\bi$ and $\bj$, these tensors are represented as the matrices
\begin{equation}
\left[\bI\right]_S = \begin{bmatrix} 1 & 0 \\ 0 & 1\end{bmatrix},\quad\quad 
\left[\bJ\right]_S=\begin{bmatrix}  0& -1 \\ 1 & 0\end{bmatrix},\quad\quad 
\left[\bK\right]_S=\begin{bmatrix} 1 & 0 \\0& -1\end{bmatrix},\quad\quad 
\left[\bL\right]_S=\begin{bmatrix} 0 & 1 \\1 & 0\end{bmatrix}
  \label{symbc} 
\end{equation}
in which form they were previously introduced by \cite{lilly18-fluids}.  Again we see the identity matrix, ninety degree counterclockwise rotation matrix, and two reflection matrices. 

Some other properties of the $\bI\bJ\bK\bL$  basis tensors will be needed.  From the transpose of an outer product, (\ref{dyadtranspose}), it follows that $\bJ^T=-\bJ$ while the other three tensors equal their own transposes.  Hence, $\bJ$ is skew-symmetric while the others are symmetric.     The identity tensor $\bI$ operates on any tensor as $\bI\bM=\bM$, while the other three tensors operate on  themselves as 
\begin{equation}
 \bK\bK=\bL\bL=\bI,\quad\quad \quad \bJ \bJ = -\bI
  \label{multrules} 
 \end{equation}
which implies that the inverse of $\bJ$ is $\bJ^T=-\bJ$ while the other three tensors are their own inverses.  The $\bJ\bK\bL$   tensors operate on each other as 
\begin{equation}
\left. \begin{array}{lllll} \displaystyle
\bJ\bK = \bL& \quad\quad& \displaystyle\bK \bL = -\bJ&\quad \quad&\displaystyle
  \bL \bJ = \bK\vspace{.1in}\\
 \displaystyle \bK\bJ = -\bL&\quad\quad&  \displaystyle\bL\bK = \bJ& \quad\quad& \displaystyle
  \bJ \bL = -\bK\\\end{array}\right.
  \label{ijklmults} 
\end{equation}
where we note that the second row is  simply the transpose of the first in view of the skew-symmetry of $\bJ$.   These combine into the easily remember triple product formulas
\begin{equation}
\bJ\bK\bL=\bK\bL\bJ=\bL\bJ\bK=\bI,\quad\quad\quad \bJ\bL\bK=\bL\bK\bJ=\bK\bJ\bL=-\bI
 \end{equation}
where the triple tensor product of $\bJ$, $\bK$, and $\bL$ is the identity tensor when these occur in alphabetical order or a cyclic permutation therefore, and the negative of the identity tensor otherwise. All of the above identities are readily shown beginning with the definitions (\ref{ijkldef}) by using the collapse rule (\ref{twodyads}) for two successive outer products, together with the fact that $\bj\cdot \bi=0$ by definition.

In addition, we have from (\ref{dyadtrace}) that $\tr\left\{\bI\right\}=2$ while the traces of the other three tensors vanish.  Any two-dimensional tensor $\bM$ can be written in terms of the  $\bI\bJ\bK\bL$ tensors in the form
\begin{equation}
\bM = \frac{1}{2}\left\{\tr\left\{\bM\right\} \bI +\tr\left\{\bM\bJ\right\} \bJ^T+\tr\left\{\bM\bK\right\} \bK+\tr\left\{\bM\bL\right\} \bL \right\}\label{Mexpand}
\end{equation}
where the tensor coefficients are expressed directly in terms of tensor traces of operations on $\bM$.  This equation is verified  by letting $\bM$ operate to the right on each of the $\bI\bJ\bK\bL$ basis tensors in turn, applying the multiplication rules (\ref{multrules}) and (\ref{ijklmults}), and noting that only $\bI$ has a nonzero trace.  

Using the $\bI\bJ\bK\bL$ basis, we can now present a decomposition of the two-dimensional gradient tensor theorem,  (\ref{flowtheoremnew}), that will illuminate its physical meaning.  For concreteness, we take $\bu$ to represent a velocity field, and  define the  usual divergence $\delta$, vorticity $\zeta$, and normal and shear strain rates $\nu$ and $\sigma$ as
\begin{equation}
\delta \equiv \frac{\partial u}{\partial x} +\frac{\partial v}{\partial y},\quad\quad
\zeta \equiv \frac{\partial v}{\partial x} -\frac{\partial u}{\partial y},\quad\quad
\nu\equiv \frac{\partial u}{\partial x} -\frac{\partial v}{\partial y},\quad\quad
\sigma \equiv \frac{\partial u}{\partial y} +\frac{\partial v}{\partial x}  \tag{\ref{dzns}}
\end{equation}
which are readily shown, via the definitions of the  $\bI\bJ\bK\bL$ tensors, to have the alternate expressions as
\begin{equation}
\delta = \bm\nabla \cdot \bu,\quad\quad
\zeta= \bm\nabla \cdot \left(\bJ^T \bu\right),\quad\quad
\nu=\bm\nabla \cdot \left(\bK \bu\right),\quad\quad
\sigma=\bm\nabla \cdot \left(\bL \bu\right). \label{divetc}
\end{equation}
These can been seen as being the divergences of the vector field $\bu$ and modified versions thereof.  We can then show that the 2D gradient tensor theorem, (\ref{flowtheoremnew}), can be decomposed into a set of four identities, one for each of the $\bI\bJ\bK\bL$ tensors,
\begin{equation}
\begin{array}{lcl}\displaystyle  \iint_\calA \displaystyle\bm\nabla \cdot \bu\, \rd A = \oint_\calL \bu \cdot\rd \bn&\quad\quad\quad\bigg|\quad\quad\quad &  \displaystyle \iint_\calA\nabla_iu^i \,\rd A =    \oint_\calL u^i \rd n_i \vspace{.1in}\\
 \displaystyle \iint_\calA  \bm\nabla \cdot \left(\bJ^T \bu\right) \rd A  = \oint_\calL \left(\bJ^T\bu\right) \cdot\rd \bn & \quad\quad\quad\bigg|\quad\quad\quad&    \displaystyle \iint_\calA   \varepsilon^{ij3}  \partial_j u_i \,\rd A = \oint_\calL\varepsilon^{ij3} u_i \,\rd n_{\! j} \vspace{.1in}\\ 
 \displaystyle \iint_\calA \bm\nabla \cdot \left(\bK \bu\right) \rd A =\oint_\calL\left(\bK\bu\right) \cdot\rd \bn&\quad\quad\quad\bigg|\quad\quad\quad&  \displaystyle \iint_\calA\nabla_i \left(K^{i}_{\cdot j}u^j\right) \,\rd A =    \oint_\calL u^i  K^{i}_{\cdot j} \,\rd n_i    \vspace{.1in}\\
 \displaystyle\iint_\calA\bm\nabla \cdot \left(\bL \bu\right)\rd A =\oint_\calL\left(\bL\bu\right) \cdot\rd \bn&\quad\quad\quad\bigg|\quad\quad\quad &    \iint_\calA\nabla_i \left(L^{i}_{\cdot j}u^j\right) \,\rd A =    \oint_\calL u^i L^{i}_{\cdot j}\,\rd n_i 
\end{array}\label{flowexpansionscalarnew}
\end{equation}
which are the divergence theorem, the Kelvin--Stokes theorem, and two related theorems involving the strain rates.   Thus, the  2D gradient tensor theorem states that the spatially integrated values of the divergence $\delta$, vorticity $\zeta$, normal strain $\nu$, and shear strain $\sigma$ are all recovered by an outer product integral along the bounding curve.  Note that the index notation expressions are here valid for general curvilinear coordinates in the horizontal plane, with the $\bK$ and $\bL$ components being specified by $K^{i}_{\cdot j} = \bg^i\bK\bg_j$ and $ L^{i}_{\cdot j} = \bg^i\bL\bg_j$.

To derive this decomposition, we expand both sides of the 2D gradient tensor theorem, (\ref{flowtheoremnew}), in terms of the $\bI\bJ\bK\bL$  basis.  The  gradient tensor $\bm\nabla \otimes \bu$ itself can be expressed as
 \begin{equation}
\bm\nabla \otimes\bu = \frac{1}{2}\left\{  \delta \bI+  \zeta\bJ^T+ \nu\bK+\sigma\bL\right\}\label{gradexpand}
\end{equation}
which may be more clear if we write out its matrix representation as
 \begin{equation}
\quad[\bm\nabla \otimes \bu]_S = \begin{bmatrix}  \frac{\partial u}{\partial x} &  \frac{\partial v}{\partial x}\vspace{.03in}\\  \frac{\partial u}{\partial y} &  \frac{\partial v}{\partial y}\end{bmatrix}=
\frac{1}{2}\left\{\delta \begin{bmatrix} 1 & 0 \\ 0 & 1\end{bmatrix} +\zeta \begin{bmatrix}  0& 1 \\ -1 & 0\end{bmatrix}+\nu\begin{bmatrix} 1 & 0 \\0& -1\end{bmatrix}+\sigma\begin{bmatrix} 0 & 1 \\1 & 0\end{bmatrix}\right\}.
\end{equation}
This is  found by substituting $\bm\nabla \otimes \bu$ for $\bM$ in (\ref{Mexpand}), then using  the fact that, as shown in the Appendix,
\begin{equation}\label{nablaresult}
\tr\left\{\left(\bm\nabla \otimes\bu\right)\bM\right\} = \bm\nabla \cdot \left(\bM^T\bu\right),
\end{equation}
then operating on $\bm\nabla\otimes \bu$ from the right by each of the $\bI\bJ\bK\bL$ tensors,  and finally employing the expressions for $\delta$, $\zeta$, $\nu$, and $\sigma$ given in  (\ref{divetc}).  If we define versions of the velocity gradient terms that are spatially averaged over the domain $\calA$, with $A$ representing its area, as
\begin{equation}
\begin{array}{ll}\displaystyle\overline\delta\equiv\frac{1}{A} \iint_\calA \bm\nabla \cdot \bu\, \rd A&\quad\quad \displaystyle
\overline\zeta\equiv\frac{1}{A} \iint_\calA  \bm\nabla \cdot \left(\bJ^T \bu\right) \rd A    \vspace{.1in}\\
 \displaystyle\overline\nu\equiv\frac{1}{A} \iint_\calA \bm\nabla \cdot \left(\bK \bu\right) \rd A&\quad\quad\displaystyle\overline\sigma\equiv\frac{1}{A}\ \iint_\calA \bm\nabla \cdot \left(\bL \bu\right) \rd A
\end{array}\label{averageequivalancediv}
\end{equation}
the the 2D gradient tensor theorem becomes 
\begin{equation}\label{otherniceexpression2}
\frac{1}{2}\left(\overline\delta\bI+\overline\zeta \bJ^T+\overline\nu\,\bK+\overline\sigma\,\bL\right)=\frac{1}{A}\oint_\calL \rd\bn \otimes\bu
\end{equation}
but we observe that the outer product on the right-hand side has the expansion
\begin{equation}
\rd\bn \otimes \bu = \frac{1}{2}\left\{\left(\bu\cdot\rd\bn \right)\bI +\left[\left(\bJ^T\bu\right)\cdot \rd \bn\right]\bJ^T+\left[\left(\bK\bu\right)\cdot \rd\bn\right]\bK+\left[\left(\bL\bu\right)\cdot\rd \bn\right]\bL \right\}\label{udn}
\end{equation}
as is found by  (\ref{Mexpand}) together with $\tr\left\{\left(\ba\otimes\bb\right)\bM\right\}  = \bb\bM\ba$,  given previously in (\ref{dyadtenstrace}), followed by the use of $\bu\bM\rd\bn=\bu\cdot  \left(\bM\rd \bn\right) = \left(\bM^T\bu\right)\cdot \rd \bn$  from the definition of the tensor transpose in (\ref{tensortranspose}). Comparing the coefficients of the basis tensors between the left-hand and right-hand sides verifies (\ref{flowexpansionscalarnew}) as claimed. For the $\bJ$ component, writing  $\left(\bJ^T\bu\right)\cdot \rd\bn =\bu \cdot \left(\bJ\rd\bn\right)=\bu \cdot \rd\bx$  shows that this is in fact Kelvin--Stokes.


In fact the expansion (\ref{flowexpansionscalarnew}) gives us another way to prove the two-dimensional gradient tensor theorem, by starting with the 2D divergence theorem.  Observe that the $\bJ^T$, $\bK$, and $\bL$ components of the gradient tensor theorem  in (\ref{flowexpansionscalarnew}) are each simply the divergence theorems for the modified flow fields $\bJ^T\bu$, $\bK\bu$, and $\bL\bu$, respectively. Thus each of these components is nothing but the divergence theorem acting on a modified velocity field, while the $\bI$ component is the divergence theorem itself; see figure~\ref{vectors} for an illustration. 

As we have seen, the two-dimensional version of the gradient tensor theorem, (\ref{flowtheoremnew}), has the divergence theorem as its trace. However, it can be written in the alternate form
\begin{equation}
\iint_\calA  \bJ \left(\bm\nabla \otimes \bu\right)\, \rd A=\oint_\calL \rd\bx \otimes\bu \label{flowtheoremv2}
\end{equation}
recalling that  $\rd\bn=\bJ^T\rd\bx$. This now resembles the Kelvin--Stokes theorem (\ref{divandstokes}b) and has that theorem as its trace, since from (\ref{dyadtrace}) we have $\tr\left\{ \bJ \left(\bm\nabla \otimes \bu\right)\right\}=\left(\bJ\bm\nabla\right)\cdot \bu=\bm\nabla \cdot \left(\bJ^T\bu\right)=\zeta$.  Since in many fluid applications the vorticity is the most important term in the velocity gradient tensor, it may be sensible to put that information along the trace, the most prominent part of the tensor.  This becomes 
\begin{equation}\label{otherniceexpression}
\frac{1}{2}\left(\overline\zeta\bI-\overline\delta \bJ^T-\overline\sigma\,\bK+\overline\nu\,\bL\right)=\frac{1}{A}\oint_\calL \rd\bx \otimes\bu
\end{equation}
and proceeding as above, we can write out the right-hand side terms to find
\begin{equation}
\overline\zeta=\frac{1}{A}\oint_\calL \bu \cdot\rd \bx,\quad\quad
\overline\delta=\frac{1}{A}\oint_\calL \left(\bJ\bu\right) \cdot\rd \bx,\quad\quad
\overline\sigma=-\frac{1}{A}\oint_\calL \left(\bK\bu \right)\cdot\rd \bx, \quad\quad
\overline\nu=\frac{1}{A}\oint_\calL \left( \bL\bu\right) \cdot\rd \bx\label{averageequivalance}
\end{equation}
for the spatially-averaged components of the gradient tensor expressed in terms of readily evaluated integrals over the  region boundary; note a minus sign only occurs in the equation for $\overline\sigma$.   These  contain equivalent information to (\ref{flowexpansionscalarnew}) in a slightly different and perhaps more readily computable arrangement.   From the definitions of the $\bI\bJ\bK\bL$ tensors, we obtain (\ref{flowexpansionscalar}) as given in the Introduction.

 \subsection{Possible extension to a curved surface*}\label{curveresults}

In this section we ask whether the two-dimensional  gradient tensor theorem also applies on a curved surface and not just on a flat surface.  Before doing so, it is useful to pursue an  alternate route to proving the 3D version.  Beginning with the left-hand side of the gradient tensor theorem in index notation, (\ref{3dtensortheoremnew}b), we write 
\begin{equation}
\iiint_\calV  \left(\nabla_i u^j\right) \bm g^i \otimes \bm g_j \rd V =\boxed{\iiint_\calV \nabla_i \left( u^j \bm g^i \otimes \bm g_j \right) \rd V = \iint_\calA  n_i u^j \bm g^i \otimes \bm g_j \rd A}\label{proof}
 \end{equation}
where the first equality follows because the covariant derivative is defined such that it vanishes for basis vectors---the metrinilic property, see (\ref{nilic})---while the second equality, in the box, is  a special case of 
\begin{equation}
\iiint_\calV \bm \nabla \cdot \mathbf{T} \rd V = \iint_\calA  \bn \cdot \mathbf{T} \rd A 
 \end{equation}
where $\bT$ is a tensor, referred to as the \emph{generalized divergence theorem} by \cite{irgens}, see (12.6.7) therein.\footnote{Note that the version of the generalized divergence theorem in \cite{irgens} differs from ours by a transpose, due to the fact that the divergence is defined therein, see (12.5.93),  to act on the last component of the tensor rather than the first.  This is another example of the transposition ambiguity discussed previously.}  

To see that the second equality in (\ref{proof}) is in fact the generalized divergence theorem, we note that it can be written  in symbolic notation as
\begin{equation}
\iiint_\calV \bm \nabla \cdot \left(\bI\otimes  \bu\right)\rd V = \iint_\calA  \bn\cdot \left(\bI\otimes \bu\right) \rd A\label{proof2}
 \end{equation}
since we have  for the left- and right-hand sides, respectively, 
\begin{align}\bm \nabla \cdot \left(\bI\otimes  \bu\right) & = \left(\bg^j  \partial_j\right) \cdot \left[\left(\bg_i \otimes \bg^i\right)\otimes \bu \right]= \bg^i\otimes \partial_i \bu = \left(\nabla_i u^j\right)\bg^i \otimes \bg_j\\  \bn \cdot \left(\bI\otimes \bu\right) &= \bn\otimes \bu = n_iu^j \bg^i \otimes  \bg_j
\end{align}
where in the first equation we have made use of the fact that $\partial_j\left(\bg_i \otimes \bg^i\right)=\Gamma_{ji}^k\bg_k\otimes \bg^i- \Gamma_{jk}^i \bg_i\otimes \bg^k=0$ due to (\ref{gammadef}).  These two quantites are in agreement with the integrands in the boxed equality in  (\ref{proof}), as claimed.  The generalized divergence theorem, in turn, is  proven by choosing the Cartesian basis to represent tensors.  With $\bT=T^{jk\ell  \ldots } \be_j\otimes\be_k\otimes\be_\ell \ldots $ being a tensor of arbitrary rank, we have
\begin{align}
 \bm\nabla \cdot \bT = \frac{\partial}{\partial x^i} \be^i  \cdot \left(T^{jk\ell  \ldots } \be_j\otimes\be_k\otimes\be_\ell \ldots  \right)&=   \left(\frac{\partial}{\partial x^i} T^{ik \ell  \ldots } \right)\be_k\otimes\be_\ell  \ldots\\
\bn \cdot \bT= n_i \be^i  \cdot \left(T^{jk\ell  \ldots } \be_j\otimes\be_k\otimes\be_\ell \ldots  \right)&=  n_i T^{ik \ell  \ldots } \be_k\otimes\be_\ell  \ldots 
 \end{align}
and the basis vectors on both sides of the equality can be pulled outside the integral, leading to 
\begin{equation}
\iiint_\calV \frac{\partial}{\partial x^i} T^{ik \ell  \ldots } \rd V = \iint_\calA n_i T^{ik \ell  \ldots }\rd A \label{expandedform}
 \end{equation}
which can be interpreted as a collection of ordinary divergence theorems for vectors given by  $T^{ik\ell \ldots}\be_i$, where the indices $k$, $\ell$, etc. act as labels specifying the coefficients of different vectors. 
 
Thus we have proven the gradient tensor theorem in two steps, first using the metrinilic property and then the generalized divergence theorem.  By representing tensors in the Cartesian basis, which is always available in Euclidean space regardless of what basis we are using otherwise, we have broken the  generalized divergence theorem into components and shown that is true by the ordinary divergence theorem.

We now show that the 2D gradient tensor theorem generalizes to two-dimensional surfaces only if they are flat. Using Greek letters to denote surfaces indices, which take on the values of 1 and 2, we can define coordinates $s^\alpha$ on the surface so that in ambient coordinates the surface is described by $q^i = q^i\left(s^1,s^2\right)$. That is, we can consider ambient coordinates restricted to the surface to be functions of the surface coordinates. Just as the ambient basis may be defined as $\bm g_i \equiv \frac{\partial }{\partial q^i} \br$ where $\br=\br(q^1,q^2,q^3)$ is the position vector considered as a function of the ambient coordinates, as in (5.2) of \cite{grinfeld}, we can define the surface basis as
\begin{equation} 
\bm s_\alpha \equiv \frac{\partial \br}{\partial s^\alpha} = \frac{\partial q^i}{\partial s^\alpha} \frac{\partial \br}{\partial q^i} = Z^i_\alpha \bm g_i, \quad\quad\quad Z^i_\alpha\equiv  \frac{\partial q^i}{\partial s^\alpha}.
\end{equation}
Here $Z^i_\alpha$, termed the \emph{shift tensor} by Grinfeld~\cite{grinfeld}, induces surface vectors from ambient vectors and can be visualized as a type of projection.  See \S10 therein for a detailed treatment of the shift tensor. 

Crucially, the surface covariant derivative is \textit{not} metrinilic with respect to the surface basis vectors. If the surface curves away from the plane spanned by the basis at some point, the derivatives of the basis vectors must be perpendicular to the surface; this is expressed in (11.16) of \cite{grinfeld}  as
\begin{equation} \nabla_\alpha \bm s_\beta = \bm n B_{\alpha\beta} 
\end{equation}
where $\bm n$ is the unit normal to the surface and  $B_{\alpha\beta}$ is called the \emph{extrinsic curvature tensor}. The eigenvalues of  $B_{\alpha\beta}$ are the principle curvatures of the surface, its trace is twice the mean curvature, and its determinant is the Gaussian curvature \cite[\S 12.4]{grinfeld}. The curvature tensor is identically zero---that is, the surface basis has vanishing covariant derivative---if and only if the embedded surface is flat. In general we have 
\begin{equation} \iint_\calA \left(\nabla_\alpha u^\beta\right) \bm s^\alpha \otimes \bm s_\beta\, \rd A \neq \iint_\calA \nabla_\alpha \left( u^\beta \bm s^\alpha \otimes \bm s_\beta \right) \rd A \label{notequal}
\end{equation}
which means that proof of the gradient tensor theorem given in (\ref{proof}) fails at its first step.  In the special case of a  flat surface, however,  we recover the gradient tensor theorem by the same logic used in the ambient space:  the equality is obtained in (\ref{notequal}),  leading to
\begin{equation} \iint_\calA \left(\nabla_\alpha u^\beta\right) \bm s^\alpha \otimes \bm s_\beta \rd A =\boxed{ \iint_\calA \nabla_\alpha \left( u^\beta \bm s^\alpha \otimes \bm s_\beta \right) \rd A = \int_\calL \bm s^\alpha \otimes \bm s_\beta \, u^\beta \rd n_\alpha}
\end{equation}
with the boxed equality being true by the generalized divergence theorem.

\section{Discussion}\label{discussion}

This paper has reintroduced the gradient tensor theorem of Gibbs~\cite{gibbs}, which seems not to be familiar to some communities that could make use of it.   At an abstract level this result can be seen as simply containing multiple cases of the fundamental theorem of calculus.  Yet the representation of a set of scalar-valued identities as a single tensor-valued identity is not just bookkeeping.  Rather, the gradient tensor theorem encodes an important understanding of the relationship between the value of the gradient of a vector-valued quantity, such as the velocity, integrated over a region and the information contained along the region's boundary.   An innovation here is the use of a tensor basis that allows the two-dimensional tensor gradient theorem to be readily applied to a region of arbitrary shape and decomposed into its constituent identities.  Because the velocity gradient tensor is a fundamental quantity in fluid mechanics, one would expect this result to find broad application to this field in particular.   A sample application to observing the ocean surface currents from a moving platform was presented.


It is natural to ask how the gradient tensor theorem is related to another generalization of the classical Kelvin--Stokes theorem, the generalized Stokes theorem of differential geometry.  As a powerful and modern framework for the calculus of potentially curved spaces in arbitrary numbers of dimensions \cite{spivak,hicks,bishop,burke,marsden}, differential geometry forms a foundational tool in general relativity \cite{carroll,needham}  that is increasingly being used in fluid dynamics as well  \cite{samelson03-mcao,cotter14-jcp,holm15-prsla,crisan22-am}.  For an orientable manifold $M$ with boundary $\partial M$, the generalized Stokes theorem states that the integral of the differential form $\omega$ over the boundary $\partial M$ is equal to the integral of the exterior derivative $\rd \omega$ over the whole of $M$
\begin{equation}
\int_M \rd \omega  = \int_{\partial M} \omega \label{generalizedstokes}
\end{equation}
see e.g. \S5--5 of Spivak \cite{spivak}. This form embodies the essence of Stokes-like integral theorems within a broad generalization of what is meant by a volume and a boundary.   To the 3D velocity field $\bu$ with Cartesian components $(u,v,w)$, one may naturally associate the 1-form $\omega_1\equiv u \rd x + v \rd y + w \rd  z$ as well as the 2-form $\omega_2\equiv w \rd y \wedge \rd z +  v \rd z \wedge \rd x +  u \rd x \wedge \rd y$.  Spivak \cite{spivak} \S5--8 \& \S5--9 shows that choosing $\omega=\omega_1$  leads to the Kelvin--Stokes theorem while choosing $\omega=\omega_2$ leads to the divergence theorem.  The remaining terms in the gradient tensor theorem are, however, not immediately recovered by the generalized Stokes theorem.  In order to incorporate them, we would need to introduce differential forms not normally considered, such as the 1-forms $u \rd x - v \rd y + w \rd z$ and $v \rd x + u \rd y + w \rd z$.  Thus while the generalized Stokes theorem does contain the tensor gradient theorem in an abstract sense, these two identities are not equivalent, and knowledge of the former certainly does not  imply an appreciation of the latter.


Finally, we note that in many applications, the two-dimensional surface of interest may be curved.   This is particularly relevant to oceanographic applications, as numerical ocean modeling and analysis of ocean observations both frequently make use of nonplanar, quasi-horizontal surfaces.    Ocean models use a variety of coordinate systems depending on the part of the ocean being modeled: one might prefer vertical or $z$-coordinates, density or $\rho$-coordinates, or bottom-following or $\sigma$-coordinates when modeling the mixed layer, the nearly adiabatic interior, or the bottom boundary layer, respectively, see e.g. \cite{griffies,fox-kemper19-fms}.   The choice of an alternate surface may be also motivated by its special properties. For example,  it is known that  transport along surfaces of constant density is orders of magnitude larger than transport across them \cite{fox-kemper13-occ},  while ``neutral surfaces'' \cite{mcdougall87-jpo,jackett97-jpo} and so-called $\bP$-surfaces \cite{nycander11-jpo} respectively allow a water parcel to  move laterally without changing its density or without requiring any work.   In attempting to apply the tensor gradient theorem to a curved surface, we found that it does not apply due to the effect of curvature terms.  However, it is possible that explicitly accounting for those terms might result in a useful generalization, or at least in boundable errors terms that would allow the theorem to be applied in an approximate sense when curvature is small.  More generally, the importance of curved surfaces in oceanography and other fields highlights the need for analysis tools that can accommodate them, and we hope that the integral theorems for flat surfaces presented herein along with a treatment of curvilinear coordinate systems  can be a step in this direction.

\appendix

\section*{Appendix}\label{appA}


This appendix  proves two equalities involving the trace, which for generality  will be shown for curvilinear coordinates using the machinery presented in \S\ref{prelim}\ref{curvilinear}.  First we consider $\tr\left\{\left(\ba\otimes\bb\right)\bM\right\}  = \bb\bM\ba$ for a spatially constant tensor $\bM$, see (\ref{dyadtenstrace}).  Expanding the left-hand side by writing  $\bM=M^i_{\cdot j} \bg_i\otimes\bg^j$ leads to
\begin{equation}
\tr\left\{\left(\ba\otimes\bb\right)\bM\right\} =
M^i_{\cdot j} \tr\left\{\left(\ba\otimes\bb\right)\left(\bg_i\otimes\bg^j\right)\right\} 
= M^i_{\cdot j} \left(\bb\cdot \bg_i\right) \left(\ba \cdot \bg^j\right) = M^i_{\cdot j}  b_i a^j 
\end{equation}
where we have made use of (\ref{dyaddyadtrace}) in the second equality.  Similarly, the right-hand side is
\begin{equation}
\bb\bM\ba = \bb\cdot \left(\bM \ba\right)=\bb \cdot \left[ M^i_{\cdot j} \left( \bg_i\otimes\bg^j  \right)\ba \right]= 
M^i_{\cdot j} \left(\bb \cdot\bg_i\right) \left(\ba \cdot\bg^j  \right)=  M^i_{\cdot j} b_i a^j
\end{equation}
and these two expressions are seen to be identical, as claimed.  Next we show a related equality for derivatives,  $\tr\left\{\left(\bm\nabla \otimes \bu\right)\bM\right\} = \bm\nabla \cdot \left(\bM^T\bu\right)$ for a spatially constant  $\bM$, as given in  (\ref{nablaresult}).  The left-hand side becomes 
\begin{equation}
\tr\left\{\left(\bm\nabla \otimes \bu\right)\bM\right\}=  \tr\left\{\left[ \left(\nabla_k u^\ell\right) \,\bg^k \otimes \bg_\ell\right]\left(M_i^{\cdot j} \bg^i\otimes\bg_j  \right) \right\}  = M_i^{\cdot j} \left(\nabla_k u^i\right) \left( \bg^k \cdot \bg_j \right) = M_i^{\cdot j} \nabla_j u^i
\end{equation}
again using (\ref{dyaddyadtrace}) together with the orthonormality condition  $\bg_i\cdot\bg^j = \delta_i^{j}$.  For the right-hand side, we have
\begin{equation}
\bm\nabla\cdot\left(\bM^T\bu\right) = \bg^k \cdot \partial_k \left(\bM^T\bu\right) = \bg^k \bM^T \partial_k\bu =\bg^k  \left(M_i^{\cdot j} \bg_j\otimes\bg^i  \right)\partial_k \bu
=M_i^{\cdot j} \bg^i \cdot \partial_j \bu = M_i^{\cdot j} \nabla_j u^i
\end{equation}
using (\ref{dyadtranspose}) for the transpose of an outer product for the third equality and (\ref{dgamma1}) for the final equality.  The preceding two expressions match, proving (\ref{nablaresult}).

\vskip6pt

\enlargethispage{20pt}

\dataccess{The model data used in this study is from a simulation called BetaEddyOne \cite{early23-zenodo}, available at Zenodo under a Creative Commons license at \url{https://doi.org/10.5281/zenodo.8200056}. }


\competing{The authors have no conflict of interests to declare.}
\funding{The work of J. Lilly, J. Feske,  and B. Fox-Kemper by grant number 2220280,  and the work of  J. Early  by grant number 2049521, both from the Physical Oceanography program of the United States National Science Foundation.}


\end{document}